\documentclass[journal,10pt]{IEEEtran}

\usepackage{cite}
\usepackage{graphicx,amssymb,amstext,amsmath,makeidx}
\usepackage{array}
\usepackage[tight,footnotesize]{subfigure}
\usepackage[normalem]{ulem}

\newtheorem{thm}{\textbf{Theorem}}

\newtheorem{cor}{\textbf{Corollary}}
\newtheorem{lem}{\textbf{Lemma}}

\newcommand{\R}{\mathbb{R}}
\newcommand{\N}{\mathbb{N}}

\newcommand{\di}{\mathrm{d}}
\newcommand{\erf}{\mathrm{erf}}
\newcommand{\erfc}{\mathrm{erfc}}
\newcommand{\E}{\mathbf{E}}
\newcommand{\Pb}{\mathbf{P}}
\newcommand{\bfx}{\mathbf{x}}
\newcommand{\bfy}{\mathbf{y}}
\newcommand{\var}{\mathrm{var}}

\newcommand{\sign}{\mathrm{sign}}
\newcommand{\dB}{\mathrm{dB}}

\renewcommand{\IEEEQED}{\IEEEQEDopen}

\begin{document}

\title{Best Signal Quality in Cellular Networks: Asymptotic Properties and Applications to Mobility Management in Small Cell Networks}

\author{Van~Minh~Nguyen,
        Fran\c{c}ois~Baccelli,
        Laurent~Thomas,
        and Chung~Shue~Chen
\thanks{Van Minh Nguyen and Laurent Thomas are with Networks and Networking Domain, Bell Labs Research, Alcatel-Lucent, Nozay, France. Email: {van\_minh}.nguyen@alcatel-lucent.com, laurent.thomas@alcatel-lucent.com.}%
\thanks{Fran\c{c}ois Baccelli and Chung Shue~Chen are with TREC, INRIA-ENS, Paris, France. Email: francois.baccelli@ens.fr, chung-shue.chen@inria.fr.}
\thanks{This work was done within the framework of the Alcatel-Lucent Bell Labs France - INRIA joint laboratory.}}

\markboth{EURASIP Journal on Wireless Communications and Networking, Special Issue on Femtocell Networks, 2010}
{Nguyen \MakeLowercase{\textit{et al.}}}

\maketitle

\begin{abstract}

    The quickly increasing data traffic and the user demand for a full coverage of mobile services anywhere and anytime are leading mobile networking into a future of small cell networks. However, due to the high-density and randomness of small cell networks, there are several technical challenges. In this paper, we investigate two critical issues: \emph{best signal quality} and \emph{mobility management}. Under the assumptions that base stations are uniformly distributed in a ring shaped region and that shadowings are lognormal, independent and identically distributed, we prove that when the number of sites in the ring tends to infinity, then (i) the maximum signal strength received at the center of the ring tends in distribution to a Gumbel distribution when properly renormalized, and (ii) it is asymptotically independent of the interference. Using these properties, we derive the distribution of the best signal quality. Furthermore, an optimized random cell scanning scheme is proposed, based on the evaluation of the optimal number of sites to be scanned for maximizing the user data throughput.

\end{abstract}


\section{Introduction}\label{sec:intro}

    Mobile cellular networks were initially designed for voice service. Nowadays, broadband multimedia services (e.g., video streaming) and data communications have been introduced into mobile wireless networks. These new applications have led to increasing traffic demand. To enhance network capacity and satisfy user demand of broadband services, it is known that reducing the cell size
is one of the most effective approaches \cite{Lee1991,Claussen2007,Chandrasekhar2008,Saunders2009}
to improve the spatial reuse of radio resources.

    Besides, from the viewpoint of end users, full coverage is particularly desirable. Although today's macro and micro cellular systems have provided high service coverage, 100\%-coverage is not yet reached because operators often have many constraints when installing large base stations and antennas. This generally results in potential coverage holes and dead zones.

    A promising architecture to cope with this problem is that of small cell networks \cite{URIE2009,Saunders2009}. A small cell only needs lightweight antennas. It helps to replace bulky roof top base stations by small boxes set on building facade, on public furniture or indoor. Small cells can even be installed by end users (e.g., femtocells). All these greatly enhance network capacity and facilitate network deployment. Pervasive small cell networks have a great potential. For example, Willcom has deployed small cell systems in Japan \cite{Chika2009}, and Vodafone has recently launched home 3G femtocell networks in the UK \cite{Judge2009}.

    In principle, high-density and randomness are the two basic characteristics of small cell networks. First, reducing cell size to increase the spatial reuse for supporting dense traffic will induce a large number of cells in the same geographical area. Secondly, end users can set up small cells by their own means \cite{Claussen2007}. This makes small cell locations and coverage areas more random and unpredictable than traditional mobile cellular networks. The above characteristics have introduced technical challenges that require new studies beyond those for macro and micro cellular networks. The main issues concern spectrum sharing and interference mitigation, mobility management, capacity analysis, and network self-organization  \cite{Chandrasekhar2008,Saunders2009}. Among these, the \emph{signal quality}, e.g., in terms of signal-to-interference-plus-noise ratio (SINR), and \emph{mobility management} are two critical issues.

    In this paper, we first conduct a detailed study on the properties of \emph{best signal quality} in mobile cellular networks. Here, the best signal quality refers to the maximum SINR received from a number of sites. Connecting the mobile to the best base station is one of the key problems. The best base station here means the base station from which the mobile receives the maximum SINR. As the radio propagation experiences random phenomena such as fading and shadowing, the best signal quality is a random quantity. Investigating its stochastic properties is of primary importance for many studies such as capacity analysis, outage analysis, neighbor cell scanning, and base station association. However, to the best of our knowledge, there is no prior art in this area.

    In exploring the properties of best signal quality, we focus on cellular networks in which the propagation attenuation of the radio signal is due to the combination of a distance-dependent path-loss and of lognormal shadowing. Consider a ring $B$ of radii $R_{\min}$ and $R_B$ such that $0 < R_{\min} < R_B < \infty$. The randomness of site locations is modeled by a uniform distribution of homogeneous density in $B$. Using extreme value theory (c.f., \cite{Leadbetter1983,Embrechts1997}), we prove that the maximum signal strength received at the center of $B$ from $n$ sites in $B$ converges in distribution to a Gumbel distribution when properly renormalized and it is asymptotically independent of the total interference, as $n \to \infty$. The distribution of the best signal quality can thus be derived.

    The second part of this paper focuses on applying the above results to mobility support in dense small cell networks. Mobility support allows one to maintain service continuity even when users are moving around while keeping efficient use of radio resources. Today's cellular network standards highlight mobile-assisted handover in which the mobile measures the pilot signal quality of neighbor cells and reports the measurement result to the network. If the signal quality from a neighbor cell is better than that of the serving cell by a handover margin, the network will initiate a handover to that cell. The neighbor measurement by mobiles is called \textit{neighbor cell scanning}. Following mobile cellular technologies, it is known that small cell networking will also use mobile-assisted handover for mobility management.

    To conduct cell scanning \cite{Nawrocki2006,WiMAX2007,TS36.3312009}, today's cellular networks use a \textit{neighbor cell list}.
    This list contains information about the pilot signal of selected handover candidates and is sent to mobiles. The mobiles then only need to measure the pilot signal quality of sites included in the neighbor cell list of its serving cell. It is known that the neighbor cell list has a significant impact on the performance of mobility management, and this has been a concern for many years in practical operations \cite{NGMN2008,NGMN2008a} as well as in scientific research \cite{Magnusson1997,Guerzoni2005,Soldani2007,Amirijoo2008}. Using neighbor cell list
is not effective for the scanning in small cell networks
    due to the aforementioned characteristics of high-density and randomness. 

The present paper proposes an optimized \emph{random cell scanning} for small cell networks. This
random cell scanning will simplify the network configuration and operation by avoiding maintaining
the conventional neighbor cell list while improving user's quality-of-service (QoS). It will
also be implementable in wideband technologies such as WiMAX and LTE.

    In the following, Section~\ref{sec:sysmodel} describes the system model. Section~\ref{sec:maxSinr} derives the asymptotic properties and the distribution of the best signal quality. Section~IV presents the optimized random cell scanning and numerical results. Finally, Section~V contains some concluding remarks.

\section{System Model}\label{sec:sysmodel}

    The underlying network is composed of cells covered by base stations with omni-directional antennas. Each base station is also called a \emph{site}. The set of sites is denoted by $\Omega \subset \N$. We now construct a model for studying the maximum signal strength, interference, and the best signal quality, after specifying essential parameters of the radio propagation and the spatial distribution of sites in the network.

    As mentioned in the introduction, the location of a small cell site is often not exactly known even to the operator. The spatial distribution of sites seen by a mobile station will hence be treated as completely random \cite{Moe2009} and will be modeled by an homogeneous Poisson point process \cite{Baccelli2009} with intensity $\lambda$.

    In the following, it is assumed that the downlink pilot signal is sent at constant power at all sites. Let $R_{\min}$ be some strictly positive real value. For any mobile user, it is assumed that the distance to his closest site is at least $R_{\min}$ and hence the path loss is the far-field. So, the signal strength of a site $i$ received by a mobile at a position $\bfy \in \R^2$ is given by
    \begin{equation}\label{eq:Pi}
        P_{i}(\bfy) = A (|\bfy - \bfx_i|)^{-\beta} X_i, \quad \textrm{for} \; |\bfy - \bfx_i| \ge R_{\min},
    \end{equation}
    where $\bfx_i \in \R^2$ is the location of site $i$, $A$ represents the base station's transmission power and the characteristics of propagation, $\beta$ is the path loss exponent (here, we consider $2 < \beta \le 4$), and the random variables $X_i = 10^{X_i^{\dB}/10}$, which represent the lognormal shadowing, are defined
from $\{X_i^{\dB}, \, i=1,2,\ldots\}$, an independent and identically distributed (i.i.d.) sequence
of Gaussian random variables with zero mean and standard deviation $\sigma_{\dB}$. Typically, $\sigma_{\dB}$ is
approximately 8~dB \cite{TR36.9422009,WiMAXForum2008}. Here, we consider that fast fading is averaged out as it varies much faster than the handover decision process.


    Cells sharing a common frequency band interfere. Each cell is assumed allocated no more than one frequency band. Denote the set of all the cells sharing frequency band $k$-th by $\Omega_k$, where $k = 1,\ldots,K$. So $\Omega_k \cap \Omega_{k'} = \emptyset$ for $k \neq k'$, and $\bigcup_{k=1}^K{\Omega_k} = \Omega$. The SINR received at $\bfy \in \R^2$ from site $i \in \Omega_k$ is expressible as
    \begin{equation}\label{eq:sinr1}
        \zeta_i(\bfy) = \frac{P_{i}(\bfy)}{N_0 + \sum_{j \neq i, j \in \Omega_k} P_{j}(\bfy)}, \quad \textrm{for} \quad i \in \Omega_k,
    \end{equation}
    where $N_0$ is the thermal noise average power which is assumed constant. For notational simplicity. Let $A := A/N_0$. Then $\zeta_i(\bfy)$ is given by
    \begin{equation}\label{eq:sinr2}
        \zeta_i(\bfy) = \frac{P_{i}(\bfy)}{1 + \sum_{j \neq i, j \in \Omega_k} P_{j}(\bfy)}, \quad \textrm{for} \quad i \in \Omega_k.
    \end{equation}
    In the following, we will use \eqref{eq:sinr2} instead of \eqref{eq:sinr1}.

\section{Best Signal Quality}\label{sec:maxSinr}

    In this section, we derive the distribution of the best signal quality. Given a set of sites $S \subset \Omega$, the \emph{best signal quality} received from $S$ at a position $\bfy \in \R^2$, denoted by $Y_S(\bfy)$, is defined as:
    \begin{equation}
        Y_S(\bfy) = \max_{i \in S} \zeta_i(\bfy).
    \end{equation}

    Let us first consider a single-frequency network (i.e., $K = 1$).

    \begin{lem}\label{thm:maxPtoSinr}
        \emph{In the cell set $S$ of single-frequency network, the site which provides a mobile the maximum signal strength will also provide this mobile the best signal quality, namely
        \begin{equation}\label{eq:maxPtoSinr}
            Y_S(\bfy) = \frac{M_S(\bfy)}{1 + I(\bfy) - M_S(\bfy)}, \quad \forall \, \bfy \in \R^2,
        \end{equation}
        where
        \begin{equation*}
            M_S(\bfy) = \max_{i \in S}{P_i(\bfy)}
        \end{equation*}
        is the maximum signal strength received at $\bfy$ from the cell set $S$, and
        \begin{equation*}
            I(\bfy) = \sum_{i \in \Omega}{P_i(\bfy)}
        \end{equation*}
        is the total interference received at $\bfy$.}
    \end{lem}
    \begin{IEEEproof} Since $\zeta_i(\bfy) = P_{i}(\bfy)/\{1 + I(\bfy) - P_i(\bfy)\}$ and $P_i(\bfy) < I(\bfy)$,
\eqref{eq:maxPtoSinr} follows from the fact that no matter which cell $i \in \Omega$ is
considered, $I(\bfy)$ is the same and from the fact that $x/(c - x)$ with $c$ constant is an increasing function of $x < c$. \end{IEEEproof}

     Let us now consider the case of multiple-frequency networks. Under the assumption that adjacent-channel interference is negligible compared to co-channel interference, cells of different frequency bands do not interfere one another. Thus, for a given network topology $\mathcal{T}$, the SINRs received from cells of different frequency bands are independent. In the context of a random distribution of sites, the SINRs received from cells of different frequency bands are therefore conditionally independent given $\mathcal{T}$. Write cell set $S$ as
    \begin{equation*}
        S = \bigcup_{k=1}^K \{S_k \, : \, S_k \subset \Omega_k\},
    \end{equation*}
with $S_k$ the subset of $S$ allocated to frequency $k$. Let
    \begin{equation*}
        Y_{S_k}(\bfy) = \max_{i \in S_k} \zeta_i(\bfy)
    \end{equation*}
    be the best signal quality received at $\bfy$ from sites which belong to $S_k$. The random variables $\{Y_{S_k}(\bfy), k=1,...,K\}$ are conditionally independent given $\mathcal{T}$. As a result,
    \begin{equation*}
        \Pb\{Y_S(\bfy) \le \gamma \, | \, \mathcal{T}\} = \prod_{k=1}^K{\Pb\{Y_{S_k}(\bfy) \le \gamma \, | \, \mathcal{T}\}}.
    \end{equation*}

    \textbf{\emph{Remark}}. For the coming discussions, we define
    \begin{equation*}
        I_S(\bfy) = \sum_{i \in S}P_i(\bfy)
     \end{equation*}
     which is the interference from cells in set $S$. In the following, for notational simplicity, the location variable $\bfy$ appearing in $Y_S(\bfy)$, $M_S(\bfy)$, $I_S(\bfy)$ and $I(\bfy)$ will be omitted in case of no ambiguity. We will simply write $Y_S$, $M_S$, $I_S$, and $I$. Note that $I_S \le I$ since $S \subset \Omega$.

    Following Lemma~\ref{thm:maxPtoSinr}, the distribution of $Y_S$ can be determined by the joint distribution of $M_S$ and $I$, which is given below.

    \begin{cor}\label{cor:exprMaxSinr}
        \emph{The tail distribution of the best signal quality received from cell set $S$ is given by
        \begin{equation}\label{eq:cdfMaxSinr1}
            \bar{F}_{Y_S}(\gamma) = \int_{u=0}^{\infty}{\int_{v=u}^{\frac{1+\gamma}{\gamma}u - 1} f_{(I,M_S)}(v,u)\di v \di u}
        \end{equation}
        where $f_{(I,M_S)}$ is the joint probability density of $I$ and $M_S$.}
    \end{cor}
    \begin{IEEEproof} By Lemma~\ref{thm:maxPtoSinr}, we have
        \begin{IEEEeqnarray*}{rCl}
            \Pb\{Y_S \geq \gamma\} & = & \Pb\big\{M_S/(1 + I - M_S) \geq \gamma \big\}\\
            & = & \Pb\big\{I \leq \frac{1 + \gamma}{\gamma}M_S - 1 \big\} \\
            & = & \int_{u=0}^{\infty}{\int_{v=u}^{\frac{1+\gamma}{\gamma}u - 1} f_{(I,M_S)}(v,u)\di v \di u}.
        \end{IEEEeqnarray*}
    \end{IEEEproof}

    In view of Corollary~\ref{cor:exprMaxSinr}, we need to study the properties of the maximum signal strength $M_S$ as well as the joint distribution of $M_S$ and $I$. As described in the introduction, in dense small cell networks, there could be a large number of neighbor cells and a mobile may thus receive from many sites with strong enough signal strength. This justifies the use of extreme value theory within this context.

    For some $R_{\min}$ and $R_B$ such that $0 < R_{\min} < R_B < \infty$, let $B \subset \R^2$ be a ring with inner and outer radii $R_{\min}$ and $R_{B}$, respectively. In this section, we will establish the following results:
    \begin{enumerate}
    \item[(i)] The signal strength $P_i$ received at the center of $B$ belongs to the \emph{maximum domain of attraction} (MDA) of the Gumbel distribution (c.f., Theorem~\ref{thm:mdaPi} in Section~\ref{sec:asymp}).
    \item[(ii)] The maximum signal strength and the interference received at the center of $B$ from $n$ sites therein are \emph{asymptotically independent} as $n \to \infty$ (c.f., Corollary~\ref{cor:asympind} in Section~\ref{sec:asymp}).
    \item[(iii)] The distribution of the best signal quality is derived (c.f., Theorem~\ref{thm:cdfMaxSinrDenseNwk} in Section~\ref{sec:cdfmaxSinr}).
    \end{enumerate}

\subsection{Asymptotic Properties}\label{sec:asymp}

    To begin with, some technical details need to be specified. Given a ring $B$ as previously defined, we will study metrics (such as e.g., signal strength, interference, etc.) as seen at the center of $B$ for a set $S \subset \Omega$ of $n$ sites located in $B$. We will use the notation $M_n$, $Y_n$, and $I_n$ instead of $M_S$, $Y_S$, and $I_S$, respectively, with
    \begin{equation*}
        M_n = \max_{i=1, i \in S}^{n}{P_i}, \; I_n = \sum_{i=1, i \in S}^{n}P_i, \; 
        Y_n = \max_{i=1, i \in S}^{n}{\zeta_i}.
    \end{equation*}

    \begin{lem}\label{lem:cdfP}
        \emph{Assume that $0 < R_{\min} < R_B < \infty$, that sites are uniformly distributed in $B$, and that the shadowing $X_i$ follows a lognormal distribution of parameters $(0, \sigma_X)$. Then the cdf of the signal strength $P_i$ received at the center of $B$ from a site located in $B$ is given by:
        \begin{multline}\label{eq:cdfPi}
            F_P(x) = c \big\{ a^{-\frac{2}{\beta}}G_1(x) - b^{-\frac{2}{\beta}}G_2(x) \\
            - e^{\nu}x^{-\frac{2}{\beta}}G_3(x) + e^{\nu}x^{-\frac{2}{\beta}}G_4(x) \big\}
        \end{multline}
        where $a = A R_{B}^{-\beta}$, $b = A R_{\min}^{-\beta}$, $c = A^{\frac{2}{\beta}} (R_{B}^2 - R_{\min}^2)^{-1}$, $\nu = 2\sigma_X^2/\beta^2$, and $G_j$, $j = 1,\dots,4$, refers to the cdf of a lognormal distribution of parameters $(\mu_j, \sigma_X)$, in which
        \begin{IEEEeqnarray*}{cCc}
            \mu_1 = \log a, & \quad & \mu_3 = \mu_1 + 2\sigma_X^2/\beta,\\
            \mu_2 = \log b, & \quad & \mu_4 = \mu_2 + 2\sigma_X^2/\beta.
        \end{IEEEeqnarray*}
        }
    \end{lem}
    \begin{IEEEproof} See Appendix~\ref{appdA}. \end{IEEEproof}

    Under the studied system model, $\{P_i, \, i = 1,2,...\}$ are independent and identically distributed (i.i.d.), and so the cdf $F_{M_n}$ and probability density function (pdf) $f_{M_n}$ of $M_n$ are directly obtained as follows:
    \begin{cor}\label{cor:distMndirect}
        \emph{Under the conditions of Lemma~\ref{lem:cdfP}, the cdf and the pdf of $M_n$ are given respectively by:
        \begin{IEEEeqnarray}{rCl}
            F_{M_n}(x) & = & F_P^n(x),             \label{eq:cdfMndirect} \\
            f_{M_n}(x) & = & n f_P(x) F_P^{n-1}(x),
            \label{eq:pdfMndirect}
        \end{IEEEeqnarray}
        where $F_P(x)$ is given by (\ref{eq:cdfPi}), and $f_P$ is the pdf of $P_i$, $f_P(x) = \di F_P(x)/\di x$.} \hfill $\square$
    \end{cor}

    Since $M_n$ is the maximum of i.i.d. random variables, we can also study its asymptotic properties by extreme value theory. Fisher and Tippett \cite[Thm.~3.2.3]{Embrechts1997} proved that under appropriate normalization, if the normalized maximum of i.i.d. random variables tends in distribution to a non-degenerate distribution $H$, then $H$ must have one of the three known forms: Fr\'echet, Weibull, or Gumbel distribution. In the following, we prove that $P_i$ belongs to the MDA of a Gumbel distribution. First of all, we establish the following result that is required to
identify the limiting distribution of $M_n$.

    \begin{lem}\label{lem:tailequivP}
        \emph{Under the conditions of Lemma~\ref{lem:cdfP}, the signal strength received at the center of $B$ from a site located in $B$ has the following tail equivalent distribution:
        \begin{subequations}
            \begin{align}
            \bar{F}_P(x) & \thicksim \kappa \frac{\exp\big(-(\log x - \mu_2)^2 / (2\sigma_X^2)\big)}{(\log x -
            \mu_2)^2/(2\sigma_X^2)} \label{eq:tailequivPowerA}\\
            & \thicksim \kappa\frac{2\sqrt{2\pi}\sigma_X\bar{G}_2(x)}{\log x - \mu_2}, \quad \textrm{as } x \to \infty,
            \label{eq:tailequivPowerB}
            \end{align}
        \end{subequations}
        where $\bar{G}_2(x) = 1 - G_2(x)$, and $\kappa = \frac{\sigma_X}{\sqrt{2\pi}\beta} \frac{R_{\min}^2}{R_{B}^2-R_{\min}^2}$.
        }
    \end{lem}

    \begin{IEEEproof} See Appendix~\ref{appdB}. \end{IEEEproof}

Equation \eqref{eq:tailequivPowerB} shows that the tail distribution of the signal strength $P_i$ is close to that of $G_2$, although it decreases more rapidly. The fact that $G_2$ determines the tail behavior of $F_P$ is in fact
reasonable, since $G_2$ is the distribution of the signal strength received from the closest possible neighboring
site (with $b = AR_{\min}^{-\beta}$ and $\sigma_X$). The main result is given below.

    \begin{thm}\label{thm:mdaPi}
        \emph{Assume that $0 < R_{\min} < R_B < \infty$, that sites are uniformly distributed in $B$, and that
shadowings are i.i.d. and follow a lognormal distribution of parameters $(0, \sigma_X)$ with $0 < \sigma_X < \infty$. Then there exists constants $c_n > 0$ and $d_n \in \R$ such that:
        \begin{equation}\label{eq:limMaxPi}
            c_n^{-1}(M_n - d_n) \overset{d}{\to} \Lambda \quad \textrm{as } n \to \infty,
        \end{equation}
        where $\Lambda$ is the standard Gumbel distribution:
        \begin{equation*}
            \Lambda(x) = \exp\{-e^{-x}\}, \quad x \in \R,
        \end{equation*}
        and $\overset{d}{\to}$ represents the convergence in distribution. A possible choice of $c_n$ and $d_n$ is:
        \begin{equation}\label{eq:maxnormconstants}
            \begin{split}
            c_n & = \sigma_X (2 \log n)^{-\frac{1}{2}} d_n,\\
            d_n & = \exp\big\{\mu_2 + \sigma_X\big(\sqrt{2 \log n} + \frac{-\log{\log n} + \log{\kappa}}{\sqrt{2 \log n}}\big)\big\},
            \end{split}
        \end{equation}
        with $\kappa$ given by Lemma~\ref{lem:tailequivP}.}
    \end{thm}

    \begin{IEEEproof} See Appendix~\ref{appdC}.\end{IEEEproof}

    \begin{figure*}[!tp]
    \centering
        \subfigure[Comparison between analytical and empirical CDFs]
        {
            \includegraphics[width=0.45\textwidth]{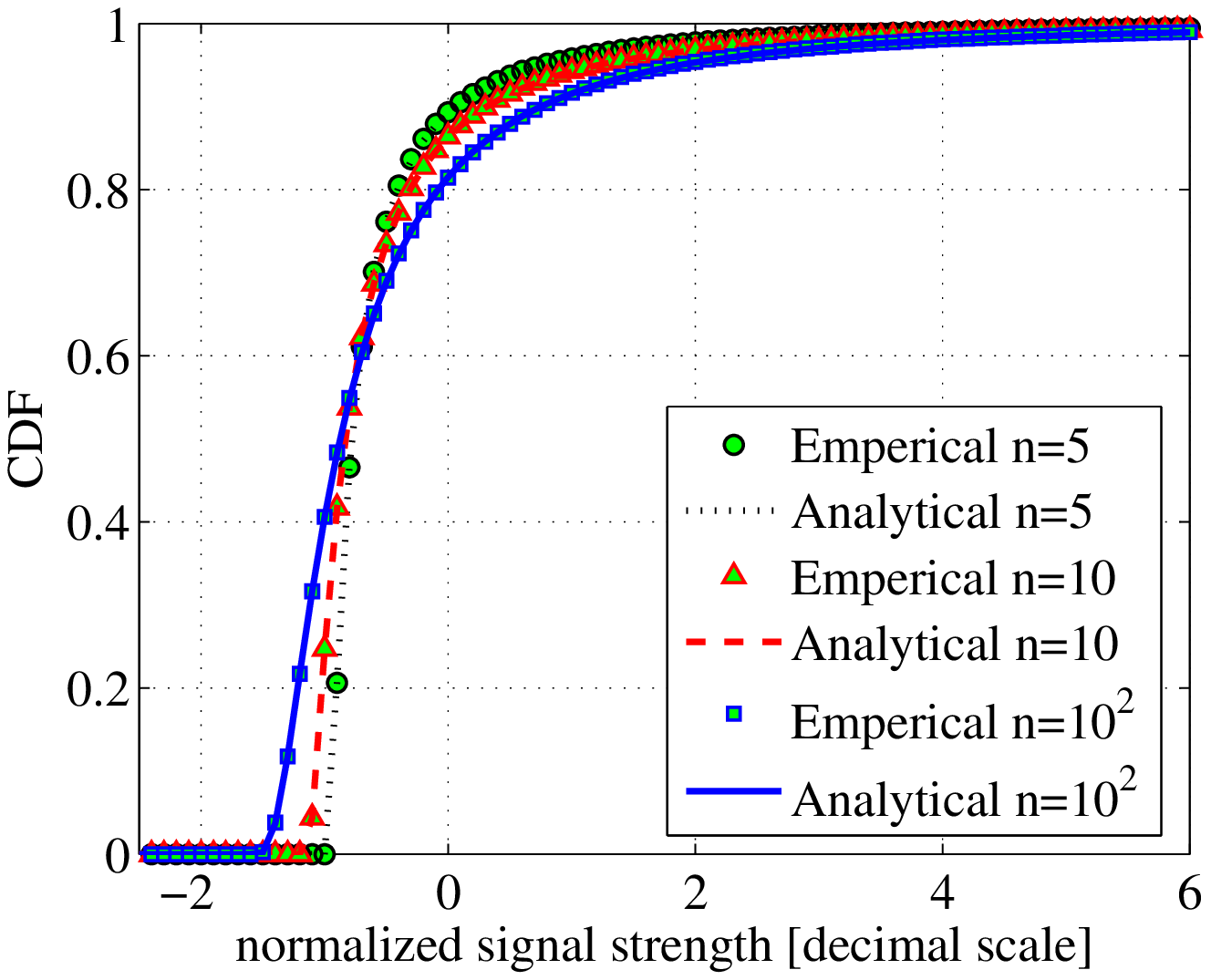}
        }
        \subfigure[Comparison between analytical and limiting CDFs]
        {
            \includegraphics[width=0.45\textwidth]{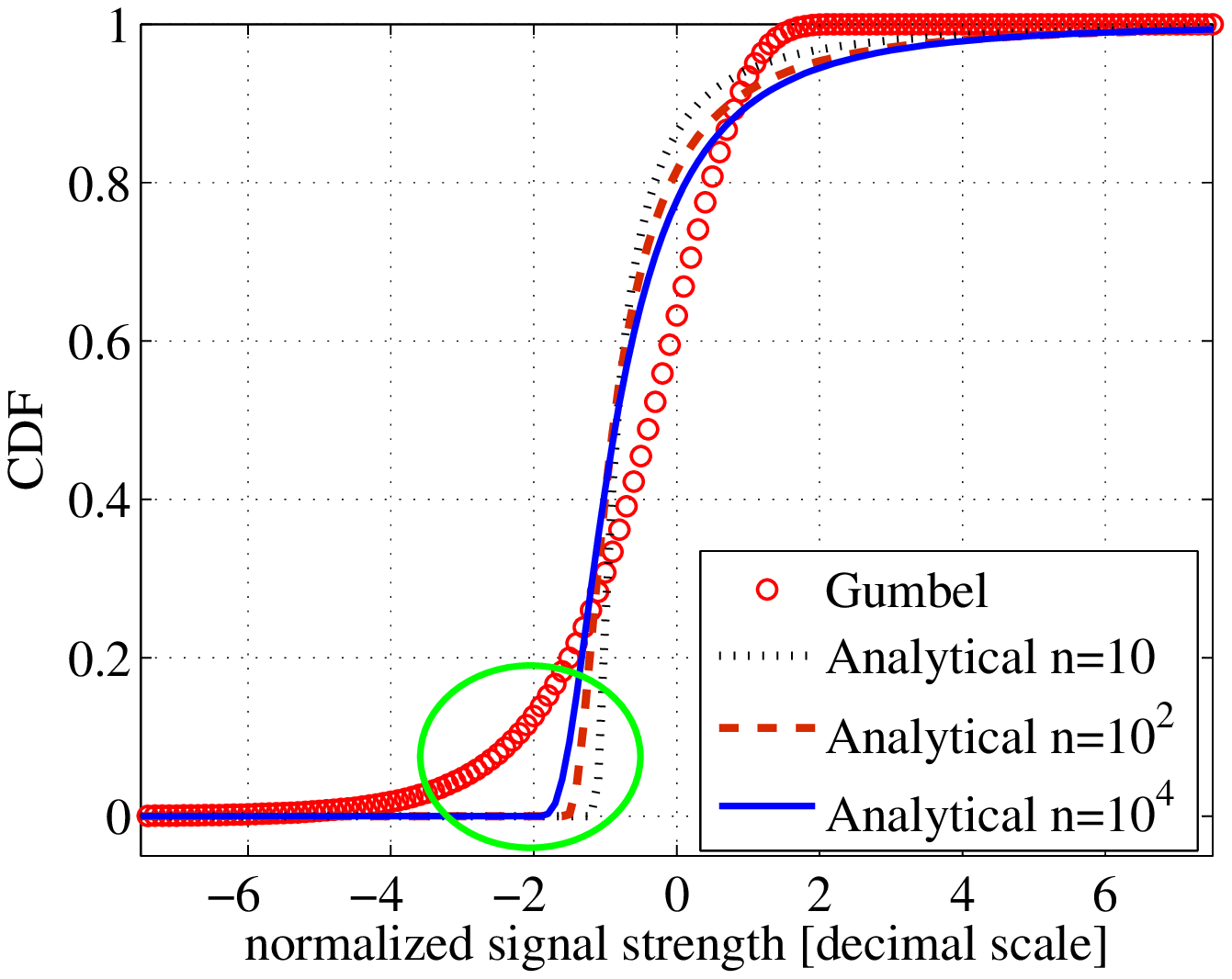}
        }
        \caption{CDF of $\tilde{M}_n$ under different $n$: $\sigma_{\dB} = 8$, $\beta = 3$.}
        \label{fig:mda_cdf}
    \end{figure*}

    \begin{figure*}[!tp]
    \centering
        \subfigure[Under different $\sigma_{\dB}$: $\beta = 3$]
        {
            \includegraphics[width=0.45\textwidth]{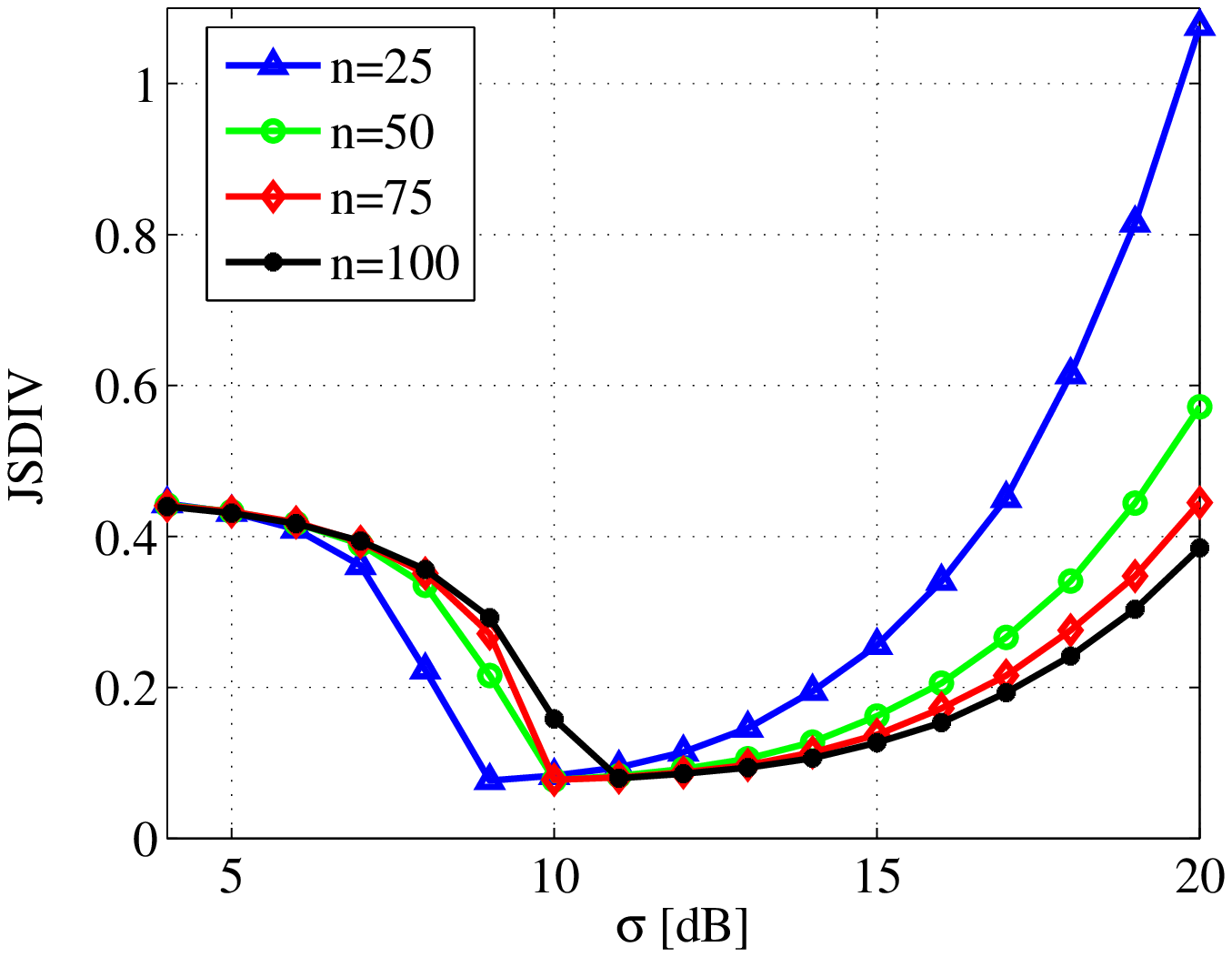}
        }
        \subfigure[Under different $\beta$: $\sigma_{\dB} = 8$]
        {
            \includegraphics[width=0.45\textwidth]{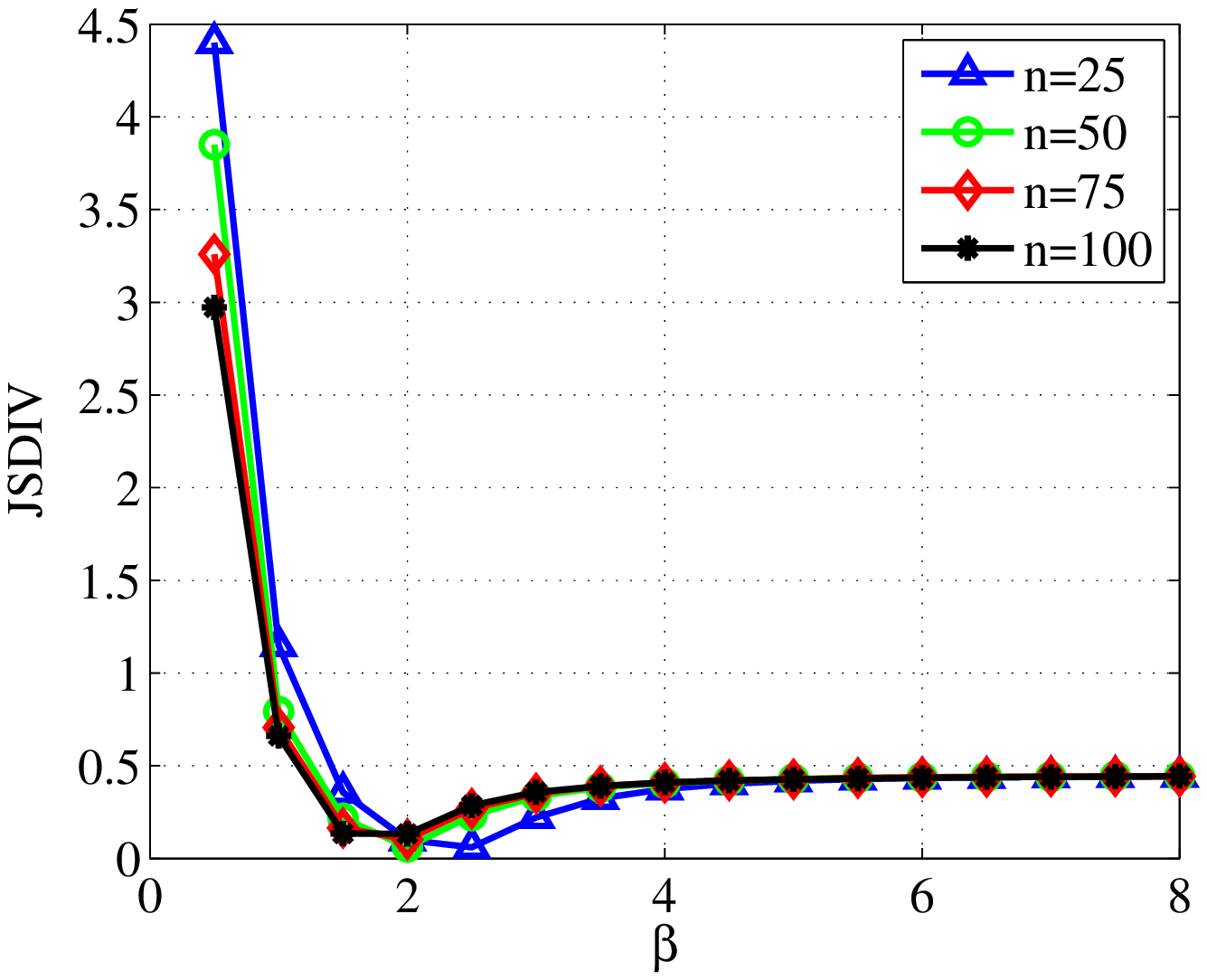}
        }\caption{Jensen-Shannon divergence between $\tilde{M}_n$ and $\Lambda$.}
        \label{fig:mda_jsdiv}
    \end{figure*}

    By Theorem~\ref{thm:mdaPi}, the signal strength belongs to the MDA of the Gumbel distribution, denoted by $\textrm{MDA}(\Lambda)$. From \cite{Chow1978,Anderson1991}, we have the following corollary of Theorem~\ref{thm:mdaPi}.

    \begin{cor}\label{cor:asympind}
        \emph{Let $\sigma_P^2$ be the variance and $\mu_P$ be the mean of signal strength $P_i$. Let $\tilde{I}_n = (I_n - n\mu_P)/(\sqrt{n}\sigma_P)$. Let $\tilde{M}_n = (M_n - d_n)/c_n$, where $c_n$ and $d_n$ are given by \eqref{eq:maxnormconstants}. Under the conditions of Theorem~\ref{thm:mdaPi},
        \begin{equation}\label{eq:convegofjoint}
            \big(\tilde{M}_n, \tilde{I}_n  \big) \! \overset{d}{\to} \big(\Lambda, \Phi\big) \quad
            \textrm{as } n \to \infty,
        \end{equation}
        where $\Lambda$ is the Gumbel distribution  and  $\Phi$ the standard Gaussian distribution,
and where the coordinates are independent.}
    \end{cor}

    \begin{IEEEproof}
        Conditions $0 < R_{\min}$ and $\sigma_X < \infty$ provide $\sigma_P^2 \le \var\{A R_{\min}^{-\beta} X_i\} < \infty$. Then the result follows by Theorem~\ref{thm:mdaPi} and \cite{Chow1978,Anderson1991}.
    \end{IEEEproof}

    Note that the total interference $I$ can be written as $I = I_n + I_n^{\textrm{c}}$ where $I_n^{\textrm{c}}$ denotes the complement of $I_n$ in $I$. Under the assumptions that the locations of sites are independent and that
shadowings are also independent, $I_n$ and $I_n^{\textrm{c}}$ are independent. The asymptotic independence between $M_n$ and $I_n$ thus induces the asymptotic independence between $M_n$ and $I$. This observation is stated in the following corollary.

    \begin{cor}\label{cor:asympindMnI}
        \emph{Under the conditions of Theorem~\ref{thm:mdaPi}, $M_n$ and $I$ are asymptotically independent as $n \to \infty$.} \hfill $\square$
    \end{cor}

    This asymptotic independence facilitates a wide range of studies involving the total interference and the maximum signal strength. This result will be used in the coming sub-section to derive the distribution of the best signal quality.


    \textbf{\emph{Remark}}:
    The asymptotic properties given by Theorem~\ref{thm:mdaPi} and Corollaries~\ref{cor:asympind} and \ref{cor:asympindMnI} hold when the number of sites in {\em a bounded area} tends to infinity. This corresponds to a network densification process in which more sites are deployed in a given geographical area in order to satisfy the need for capacity, which is precisely the small cell setting.

    \begin{figure*}[!t]
        \centering
        {
            \includegraphics[width=0.9\textwidth]{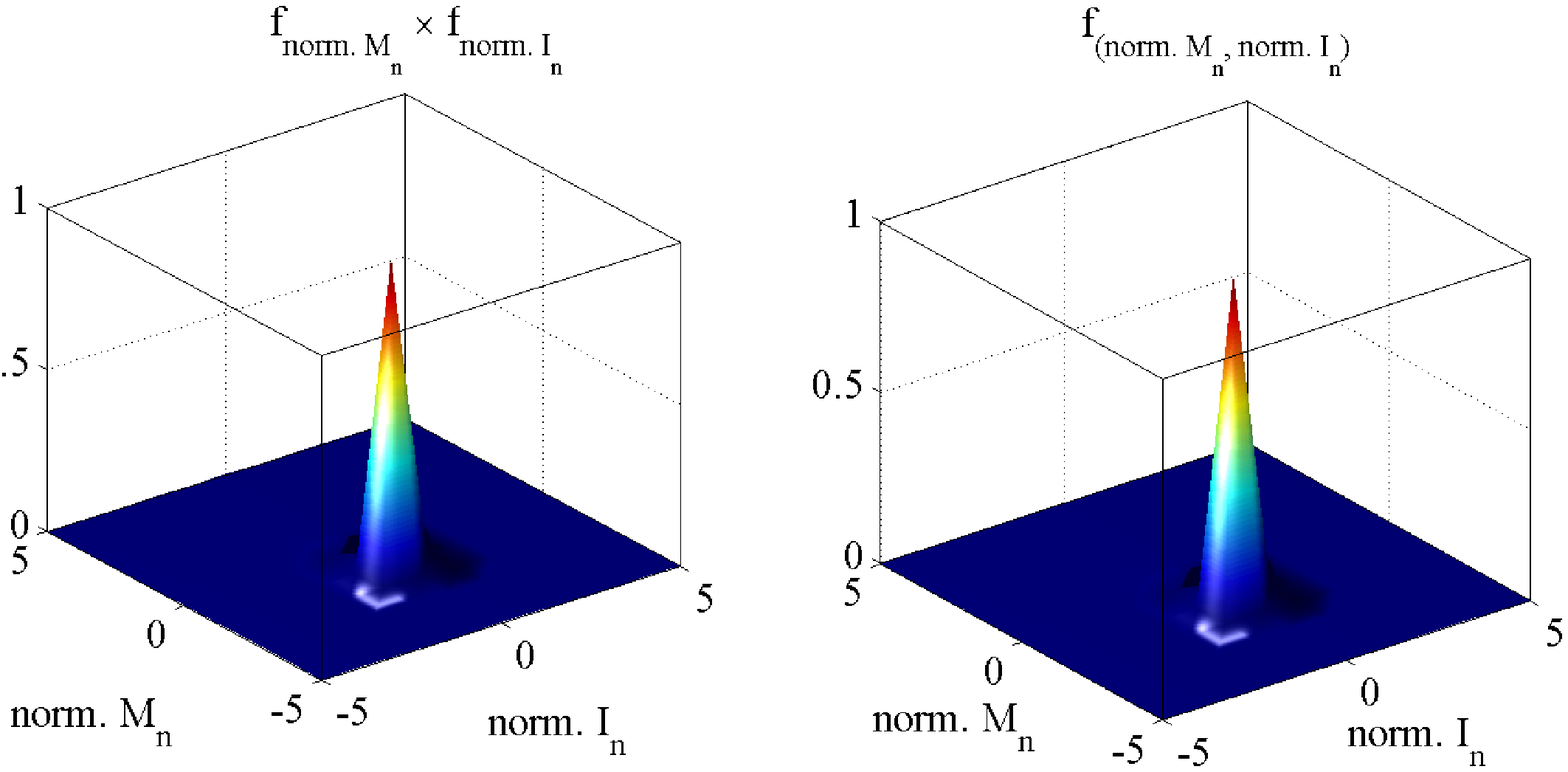}
        }
        \caption{Example of joint densities of $\tilde{M}_n$ and $\tilde{I}_n$: $n=50$, $\sigma_{\dB} = 8$, $\beta = 3$. Here,  norm.~$M_n$ refers to $\tilde{M}_n$, while norm.~$I_n$ refers to $\tilde{I}_n$.}
        \label{fig:asympind_pdf}
    \end{figure*}
    \begin{figure*}[!tp]
    \centering
        \subfigure[Under different $\sigma_{\dB}$: $\beta = 3$]
        {
            \includegraphics[width=0.45\textwidth]{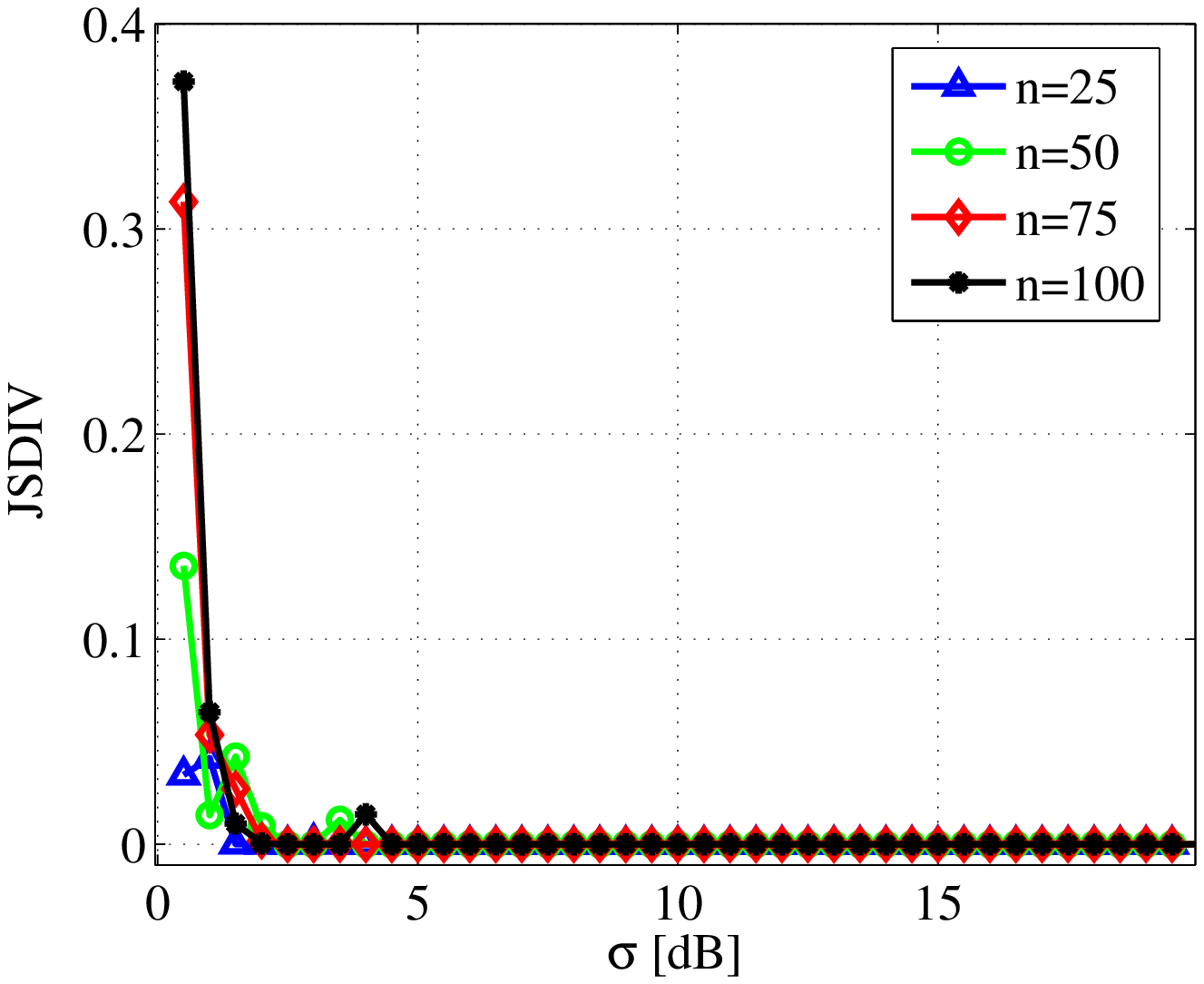}
        }
        \subfigure[Under different $\beta$: $\sigma_{\dB} = 8$]
        {
            \includegraphics[width=0.45\textwidth]{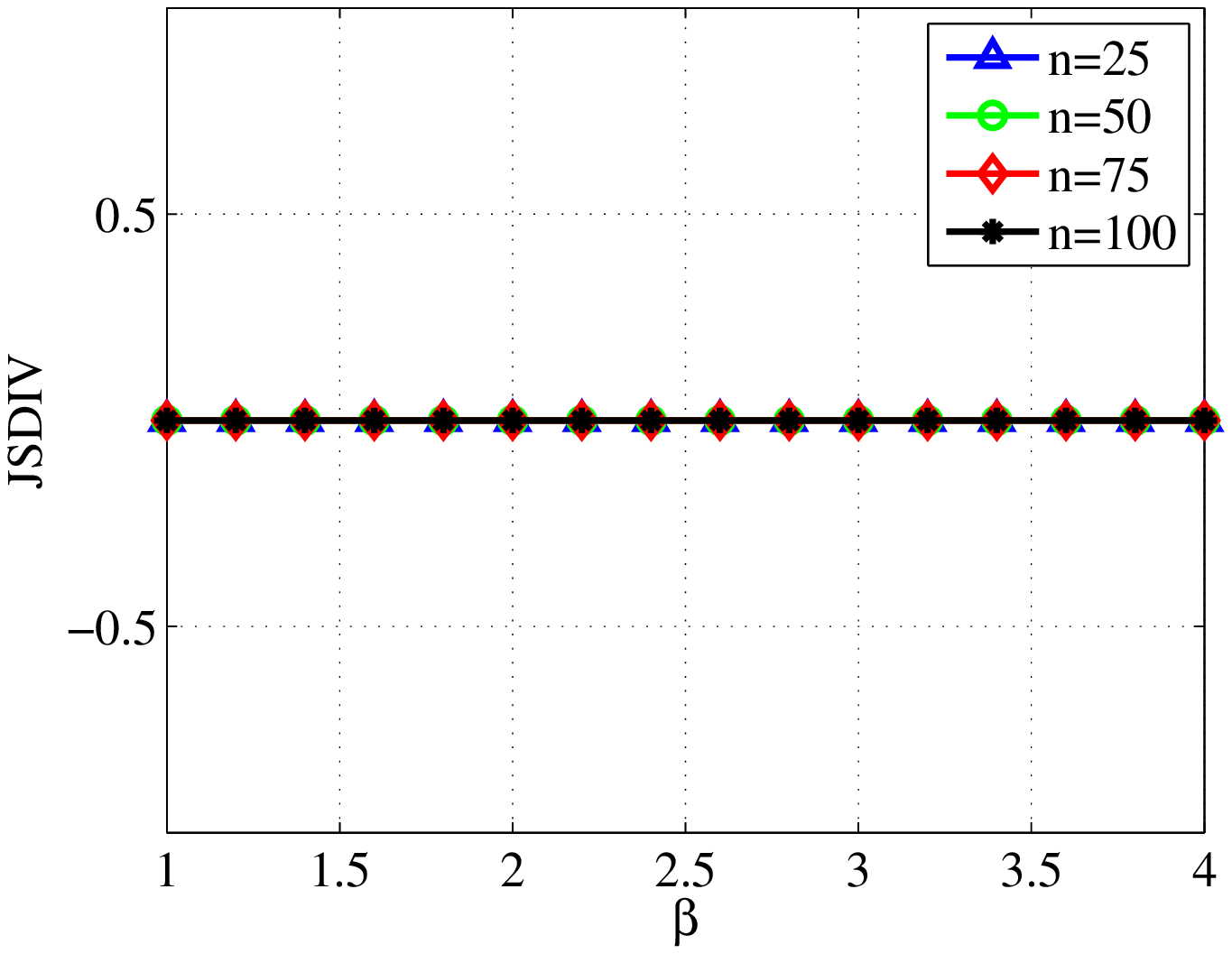}
        }
        \caption{Jensen-Shannon divergence between $f_{(\tilde{M}_n, \tilde{I}_n)}$ and $f_{\tilde{M}_n} \times f_{\tilde{I}_n}$.}
        \label{fig:asympind_jsdiv}
    \end{figure*}

\subsection{Convergence Speed of Asymptotic Limits}

    Theorem~\ref{thm:mdaPi} and Corollaries~\ref{cor:asympind} and \ref{cor:asympindMnI} provide asymptotic
properties when $n \to \infty$. In practice, $n$ is the number of cells to be scanned,
and so it can only take moderate values. Thus, it is important to evaluate the speed of convergence speeds of \eqref{eq:limMaxPi} and \eqref{eq:convegofjoint}. We will do this based on simulations and will measure the discrepancy
using a symmetrized version of the Kullback-Leibler divergence (the so called Jensen-Shannon divergence (JSdiv)).

    Let us start with some numerical evaluations of the convergence of $\tilde{M}_n$ to its limiting distribution. Fig.~\ref{fig:mda_cdf}(a) shows $F_{\tilde{M}_n}$ for different $n$ and compares to empirical simulation results.
As expected the analytical distributions obtained by (\ref{eq:cdfMndirect}) of
Corollary~\ref{cor:distMndirect} match with the empirical distributions for all $n$.
Fig.~\ref{fig:mda_cdf}(b) plots the analytical distribution and its limiting distribution, i.e. the
Gumbel distribution $\Lambda$. There is a discrepancy in the negative regime (see the circled region
in Fig.\ref{fig:mda_cdf}(b)). It is worth pointing out that as a maximum of signal strengths, $M_n \ge 0$
and thus $\tilde{M}_n \ge - d_n/c_n = -\sqrt{2\log{n}}/\sigma_X$ since $\tilde{M}_n = (M_n - d_n)/c_n$.
This means that $F_{\tilde{M}_n}(x) = 0$,  $\forall x \le  -\sqrt{2\log{n}}/\sigma_X$,
whereas $\Lambda(x) > 0$, $\forall x > -\infty$. This explains the gap observed for $n$ small.
This dissimilarity should have limited impact as long as we only deal with positive values of $M_n$
(respectively, $\tilde{M}_n \geq -\sqrt{2\log{n}}/\sigma_X$).

    We now study the symmetrized divergence between the analytical and limiting distributions
of $\tilde{M}_n$ for some moderate values of $n$ and under different $\sigma_{\dB}$ and $\beta$.
The convergence is best for $\sigma_{\dB}$ around 10~dB and $\beta$ around two to four.
For practical systems, $\sigma_{\dB}$ is approx. 8~dB and $2 < \beta \leq 4$.
We compute the Jensen-Shannon divergence for $\beta = 3$ and $\sigma_{\dB} = 8$ and plot
the results in Fig.~\ref{fig:mda_jsdiv}(a) and (b), respectively.
For these (and other) values (within the range given above) of
$\sigma_{\dB}$ and $\beta$, $\tilde{M}_n$ and $\Lambda$ have low divergence.

    Let us now measure the (dis)similarity between the empirical joint distribution,
$\Pb\{\tilde{M}_n \leq u, \tilde{\mathrm{I}}_n \leq v\}$, and the product of the empirical
marginal distributions, $\Pb\{\tilde{M}_n \leq u\} \times \Pb\{\tilde{\mathrm{I}}_n \leq v \}$.
Fig.~\ref{fig:asympind_pdf} shows an example with $n = 50$, $\beta = 3$ and $\sigma_{\dB} = 8$.
We see that these two density functions are very similar.
Fig.~\ref{fig:asympind_jsdiv} compares these two density functions for different
values of $\sigma_{\dB}$ and $\beta$. Within the range defined above,
the divergence between the two distributions is again small.
Comparing Fig.~\ref{fig:mda_jsdiv} and Fig.~\ref{fig:asympind_jsdiv}, one can conclude
that even if the convergence of $\tilde{M}_n \overset{d}{\to} \Lambda$ remains slow,
$\tilde{M}_n$ and $\tilde{I}_n$ quickly become uncorrelated. Thus, the independence between
${M}_n$ and ${I}_n$ holds for moderate values of $n$, i.e.,
    \begin{equation}\label{eq:asympjointIandM}
        f_{(M_n,I_n)}(u,v) \approx f_{M_n}(u) \times f_{I_n}(v),
    \end{equation}
and so the same conclusion holds for the independence between ${M}_n$ and $I$.

\subsection{Distribution of the Best Signal Quality}\label{sec:cdfmaxSinr}


    From the above results, we have the distribution of $M_n$ and the asymptotic independence between $M_n$ and $I$. In order to derive the distribution of the best signal quality, we also need the distribution of the total interference.
    \begin{lem}\label{lem:cfI}
        \emph{Assume that shadowings are i.i.d. and follow a lognormal distribution of
parameters $(0, \sigma_X)$, $\beta > 2$, and that sites are distributed according to a
Poisson point process with intensity $\lambda$ outside the disk of radius $R_{\min} > 0$.
Let $I$ be the interference received at the disk center, and $\phi_I$ be the characteristic
function of $I$. Then:
\begin{enumerate}
\item \begin{equation}\label{eq:cfI:exact}
\hspace{-.8cm} \phi_I(w) = \exp\bigr\{- \pi\lambda\alpha(A|w|)^{\alpha} \int_0^{\frac{A|w|}{R_{\min}^{\beta}}}\frac{1 - \phi_X(\sign(w)t)}{t^{\alpha + 1}} \di t\bigr\},
\end{equation}
where $\alpha = 2/\beta$, and $\phi_X$ is the characteristic function of $X_i$.
          \item $|\phi_I(w)|^p \in \mathbb{L}$ for all $p = 1, 2, \ldots$, where $\mathbb{L}$ is the space of absolutely integrable functions.
          \item If $A R_{\min}^{-\beta}$ is large, then $\phi_I$ admits the following approximation:
              \begin{equation}\label{eq:cfI:approx}
                \phi_{I}(w) \approx \exp\big(- \delta |w|^{\alpha} [1 - j \sign(w)\tan(\frac{\pi\alpha}{2})] \big),
              \end{equation}
              where $\delta = \pi\lambda A^{\alpha} \exp(\frac{1}{2}\alpha^2 \sigma_X^2) \Gamma(1-\alpha)\cos(\pi\alpha/2)$, with $\Gamma(\cdot)$ denoting the gamma function.
        \end{enumerate}
        }
    \end{lem}
    \begin{IEEEproof} See Appendix~\ref{appd:lem:cfI}. \end{IEEEproof}

    \begin{thm}\label{thm:cdfMaxSinrDenseNwk}
        \emph{Under the assumptions of Lemma~\ref{lem:cfI}, let $B$ be the ring of inner and outer radii $R_{\min}$ and $R_B$, respectively. Denote the best signal quality received at the center of $B$ from $n$ sites in $B$ by $Y_n$. Assume that $0 < \sigma_X < \infty$, $0 < R_{\min} < R_B < \infty$, that $A R_{\min}^{-\beta}$ is large,
and that $\pi \lambda (R_{B}^2 - R_{\min}^2) > n$, with high probability, where $n$ is some
positive integer. Then the tail distribution of $Y_n$ can be approximated by:
        \begin{multline}\label{eq:cdfMaxSinrDenseNwk}
            \!\!\!\! \bar{F}_{Y_n}(\gamma) \approx \int_{\gamma}^{\infty}\!\!\!\Big\{f_{M_n}(u) \!\!\!
            \int_{0}^{\infty}\frac{2 }{\pi w}e^{-\delta w^{\alpha}} \sin\big(w\frac{u-\gamma}{2\gamma}\big) \\
            \times \cos\big(w u + w\frac{u-\gamma}{2\gamma} - \delta w^{\alpha}\tan{\frac{\pi \alpha}{2}}\big) \di w \Big\}\di u.
        \end{multline}
        }
    \end{thm}

    \begin{IEEEproof}
        See Appendix~\ref{appdD}.
    \end{IEEEproof}

    The approximation proposed in Theorem 2 will be used in Section~\ref{sec:rndscan} below. It will be validated by simulation in the context considered there.

\section{Random Cell Scanning for Data Services}\label{sec:rndscan}

    In this section, the theoretical results developed in Section~\ref{sec:maxSinr} are applied to random cell scanning.

\subsection{Random Cell Scanning}

    Wideband technologies such as WiMAX, WCDMA, and LTE use a predefined set of codes for the identification of cells at the air interface. For example, 114 pseudo-noise sequences are used in WiMAX \cite{IEEE802.162009}, while 504 physical cell identifiers are used in LTE \cite{TS36.3002009}. When the mobile knows the identification code of a cell, it can synchronize over the air interface and then measure the pilot signal quality of the cell. Therefore, by using a predefined set of codes, these wideband technologies can have more autonomous cell measurement conducted by the mobile. In this paper, this identification code is referred to as cell synchronization identifier (CSID).

    In a dense small cell network where a large number of cells are deployed in the same geographical area, the mobile can scan any cell as long as the set of CSIDs used in the network is provided. This capability motivates us to propose \emph{random cell scanning} which is easy to implement and has only very few operation requirements. The scheme is detailed below:
    \begin{enumerate}
      \item[(1)] When a mobile gets admitted to the network, its (first) serving cell provides him the whole set of CSIDs used in the network. The mobile then keeps this information in its memory.
      \item[(2)] To find a handover target, the mobile randomly selects a set of $n$ CSIDs from its memory and conducts the standardized scanning procedure of the underlying cellular technology, e.g., scanning specified in IEEE 802.16 \cite{IEEE802.162009}, or neighbor measurement procedure specified in 3G \cite{TS25.3312009} and LTE \cite{TS36.3312009}.
      \item[(3)] The mobile finally selects the cell with the best received signal quality as the handover target.
    \end{enumerate}

    In the following, we determine the number of cells to be scanned which maximizes the data throughput.

\subsection{Problem Formulation}\label{sec:pblformulation}

    The optimization problem has to take into account the two contrary effects due to the number of cells to be scanned. On one hand, the larger the set of scanned cells, the better the signal quality of the chosen site,
and hence the larger the data throughput obtained by the mobile. On the other hand, scanning can have a linear
cost in the number of scanned cells, which is detrimental to the throughput obtained by the mobile.

Let us quantify this using the tools of the previous sections.

    Let $W$ be the average cell bandwidth available per mobile and assume that it is a constant.
Under the assumption of additive white Gaussian noise, the maximum capacity $\xi_n$ that the
mobile can have by selecting the best among $n$ randomly scanned cells is
    \begin{equation}\label{eq:shannon}
        \xi_n = W\log(1 + Y_n).
    \end{equation}
Hence
    \begin{IEEEeqnarray*}{rCl}
        \E\{\xi_n\} & = & W\E\{\log(1 + Y_n)\} \\
        & = & W \int_{\gamma=0}^{\infty}\log(1+\gamma)f_{Y_n}(\gamma)\di\gamma,
    \end{IEEEeqnarray*}
     where $f_{Y_n}$ is the pdf of $Y_n$. By an integration by parts of $\log(1+\gamma)$ and $f_{Y_n}(\gamma)\di\gamma = -\di\bar{F}_{Y_n}(\gamma)$, this becomes:
    \begin{equation}
        \E\{\xi_n\} = W\int_{\gamma=0}^{\infty}\frac{\bar{F}_{Y_n}(\gamma)}{1+\gamma}\di\gamma.
     \end{equation}

    Note that $\E\{\xi_n\}$ is the expected throughput from the best cell.
Since $Y_n$ is the maximum signal quality of the $n$ cells, $Y_n$ increases with $n$ and so does $\xi_n$.
Hence, the mobile should scan as many cells as possible. However, on the other hand, if scanning many cells,
the mobile will consume much time in scanning and thus have less time for data transmission with the serving
cell. A typical situation is that where the scanning time increases proportionally with the number of cells
scanned and where the data transmission is suspended. This for instance happens if the underlying cellular
technology uses a \emph{compressed mode} scanning\footnote{In this mode, scanning intervals,
where the mobile temporarily suspends data transmission for scanning neighbor cells,
are interleaved with intervals where data transmission with the serving cell is resumed.}, like
e.g., in IEEE~802.16 \cite{IEEE802.162009} and also inter-frequency cell measurements defined by
3GPP \cite{TS25.3312009,TS36.3312009}.

Another scenario is that of
\textit{parallel scanning-transmission}: here scanning can be performed
in parallel to data transmission so that no transmission gap occurs; this is the
case in e.g., intra-frequency cell measurements in WCDMA \cite{TS25.3312009} and LTE \cite{TS36.3312009}.

    Let $T$ be the average time during which the mobile stays in the tagged cell
and receives data from it. Let $s$ be the time needed to scan one cell
(e.g., in WCDMA, the mobile needs $s = 25~\textrm{ms}$ if the cell is in the neighbor
cell list and $s = 800~\textrm{ms}$ if not \cite{TS25.1332009}, whereas in WiMAX, $s = 10~\textrm{ms}$, i.e.,
two 5-ms frames). Let $L(n)$ be the duration of the suspension of data transmission due
to the scanning of the $n$ cells:
    \begin{equation}
        L(n) = \begin{cases} s \times n & \textrm{if \emph{compressed mode} is used,} \\
        0 & \textrm{if \emph{parallel scan.-trans.} is enabled.} \end{cases}
    \end{equation}
Finally, let $\E\{\xi_0\}$ be the average throughput received from the serving cell
when no scanning at all is performed (this would be the case if the mobile would
pick as serving site one of the sites of set $S$ at random).

    The gain of scanning $n$ cells can be quantified by
the following metric, that we will call the \emph{acceleration}:
    \begin{eqnarray}\label{eq:productivity}
        \rho_n & \triangleq &\frac{T \cdot \E\{\xi_n\}}{T \cdot \E\{\xi_0\} + L(n) \cdot \E\{\xi_0\}} \nonumber \\
        & = & \frac{T}{T + L(n)}\times \frac{\E\{\xi_n\}}{\E\{\xi_0\}}.
    \end{eqnarray}
In this definition, $T \cdot \E\{\xi_n\}$ (resp. $T \cdot \E\{\xi_0\} + L(n) \cdot \E\{\xi_0\})$)
is the expected amount of data transmitted when scanning $n$ cells (resp. doing no scanning at all).
We aim at finding the value of $n$ that maximizes the acceleration $\rho_n$.

It is clear that $T/(T + L(n)) = 1$ when (i) $T \to \infty$, i.e., the mobile stays in and receives data
from the tagged cell forever, or (ii) $L(n) = 0$, i.e., parallel scanning-transmission is enabled.
In these cases, $\rho_n$ increases with $n$ and the mobile ``should'' scan as many cells as possible.
However, $\rho_n$ is often concave and the reward
of scanning then decreases. To characterize this, we introduce a
\emph{growth factor} $g$ defined as follows:
    \begin{equation}\label{eq:growth}
        g_n \triangleq \frac{\rho_n}{\rho_{n-1}} = \frac{T + L(n-1)}{T + L(n)} \times \frac{\E\{\xi_n\}}{\E\{\xi_{n-1}\}}.
    \end{equation}

    \begin{figure*}[!t]
    \centering
        \subfigure[Plot of $\frac{\E\{\xi_n\}}{\max_k \E\{\xi_k\}}$]
        {
            \includegraphics[width=0.45\textwidth]{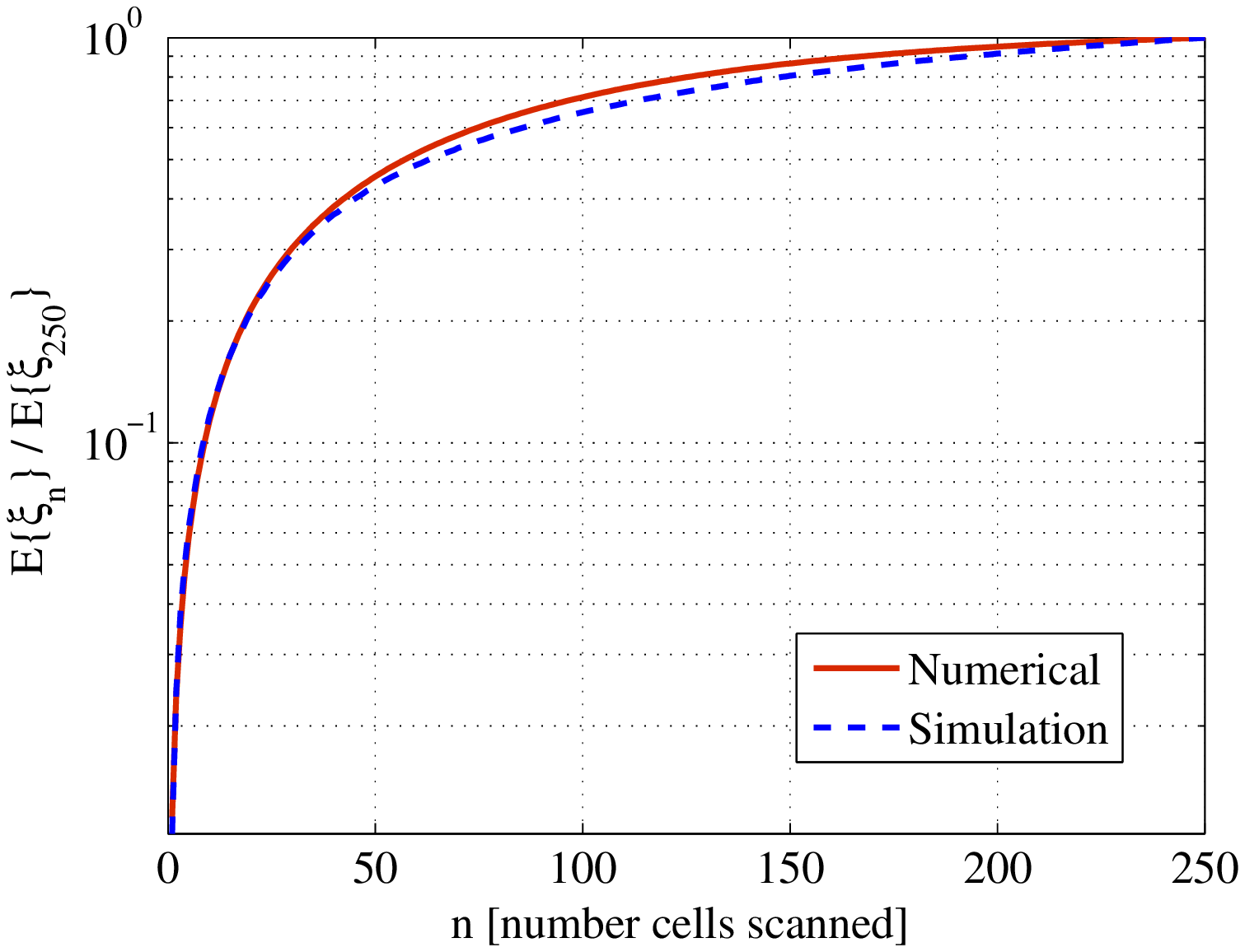}
        }
        \subfigure[Plot of $\frac{\rho_n}{\max_k \rho_k}$, $T = 0.5$~second]
        {
            \includegraphics[width=0.45\textwidth]{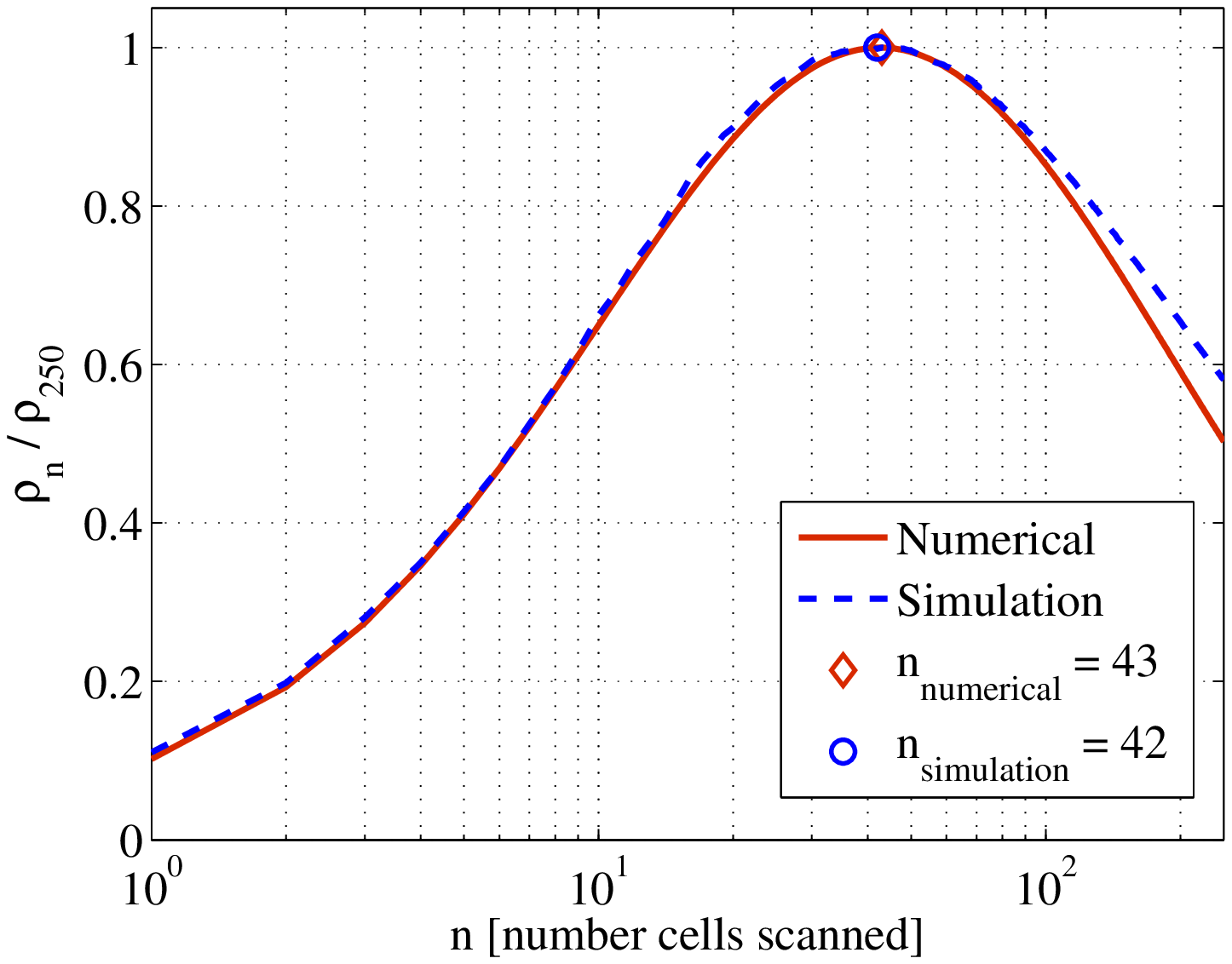}
        }\\
        \subfigure[Optimal number of cells to be scanned]
        {
            \includegraphics[width=0.45\textwidth]{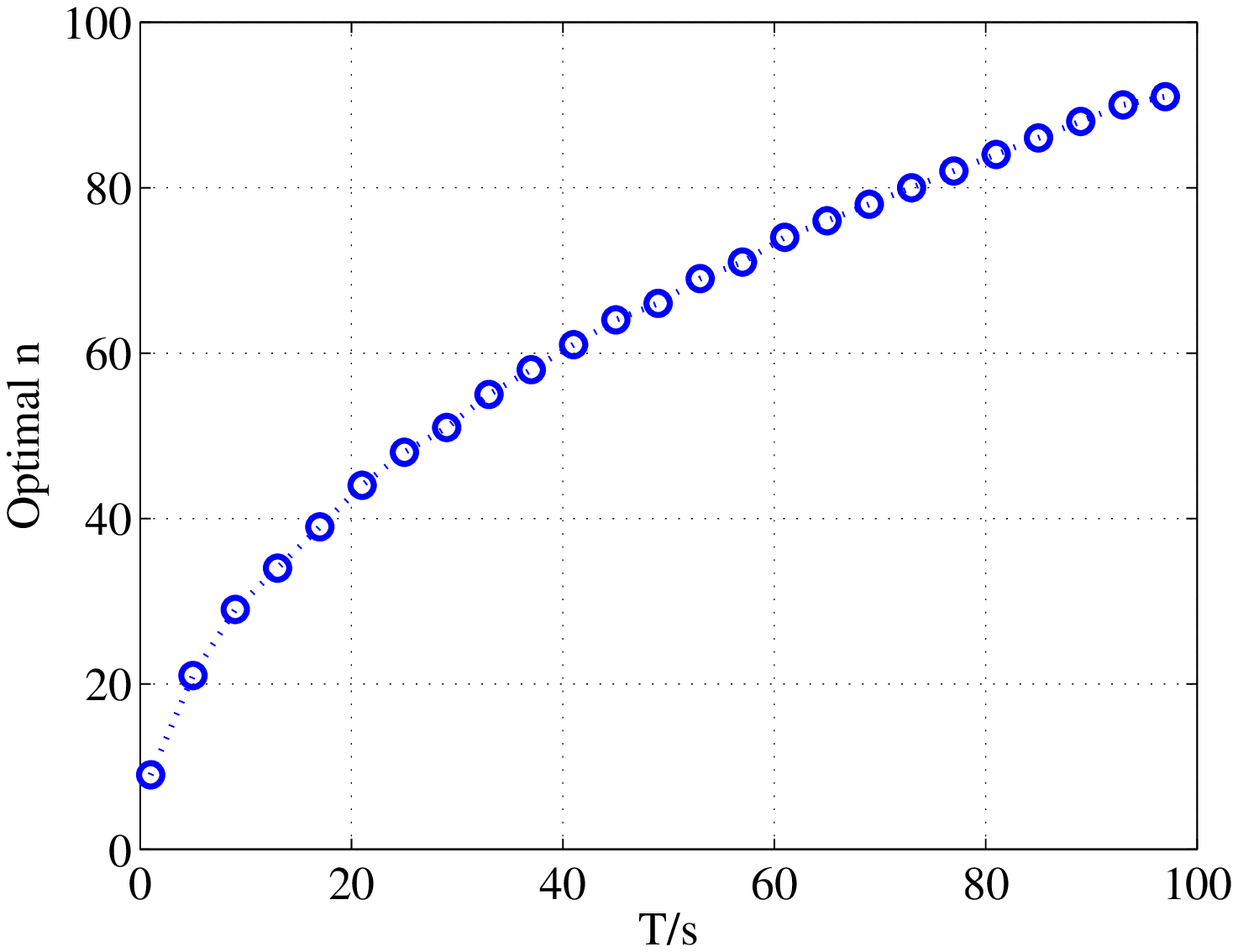}
        }
        \subfigure[Growth factor $g_n$ under different $T$]
        {
            \includegraphics[width=0.45\textwidth]{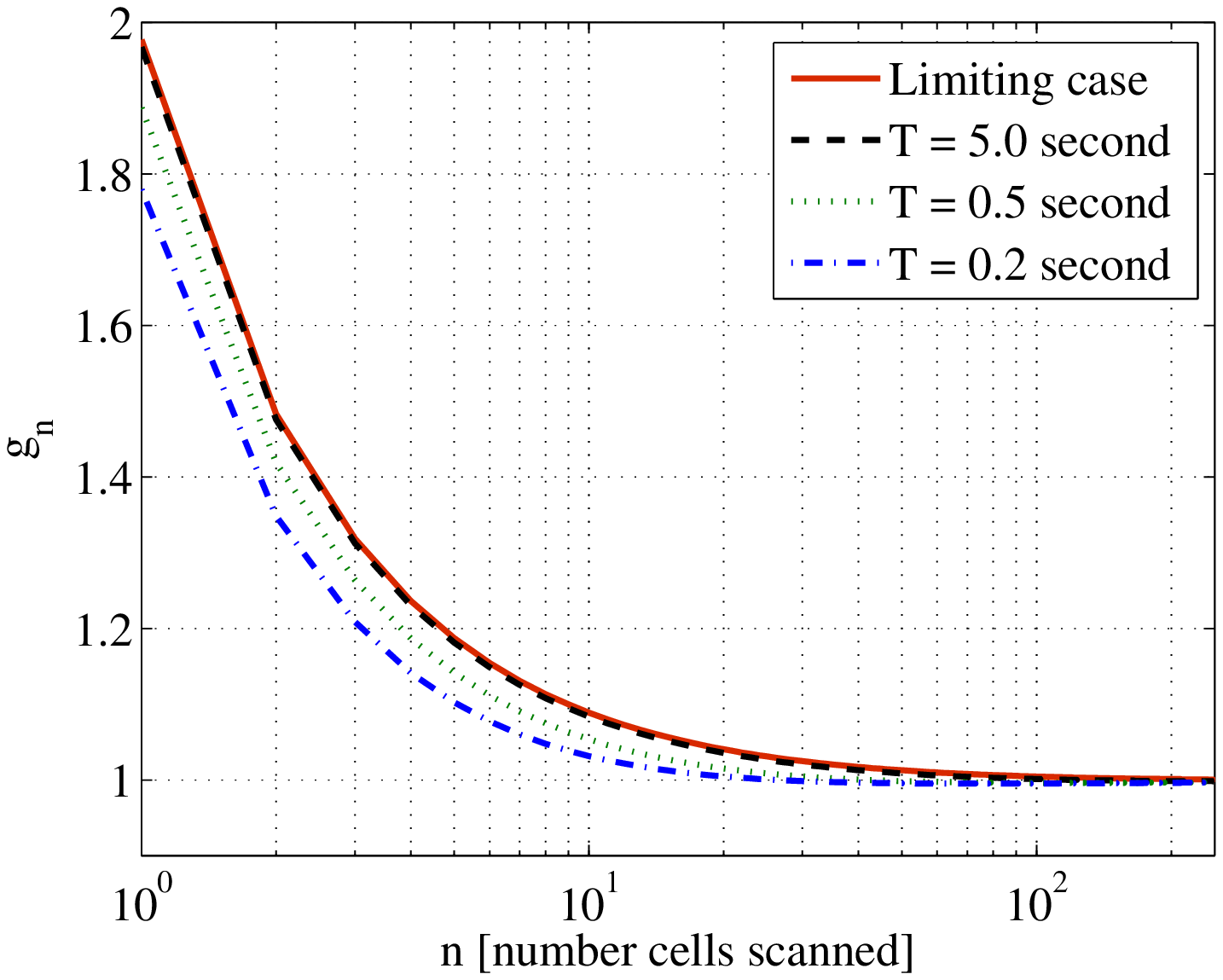}
        }
        \caption{Numerical results in the random cell scanning optimization.}\label{fig:numresult}
    \end{figure*}


    Special cases as those considered above can be cast within a general framework which consists
in finding the value of $n$ that maximizes $\rho_n $ under the constraint that
$ g_n \geq 1 + \Delta_g$, where $\Delta_g > 0$ is a threshold.

\subsection{Numerical Result}\label{sec:numeval}

    In the following, we show how to apply the above results to find the optimal $n$. We adopt WCDMA as the underlying cellular network technology. 100 omni-directional small cell base stations are deployed in a square domain of $1~\textrm{km} \times 1~\textrm{km}$. The network density is thus equal to:
    \begin{equation*}
        \lambda = 10^{-4}~\textrm{small base stations}/\textrm{m}^2.
    \end{equation*}

    It is assumed that any cell synchronization identifier can be found in a radius $R_B = 1~{\textrm{km}}$. We take  $R_{\min}$ equal to 2 meters. The propagation path loss is modeled by the picocell path loss model \cite{TR101.1121998}:
    \begin{equation}
        PL_{[\dB]}(d) = 37 + 30\log_{10}(d) + 18.3 f^{(\frac{f+2}{f+1}-0.46)},
        \label{eq:PL_dB}
    \end{equation}
    where $d$ is the distance from the base station in meters, $f$ the number of penetrated floors in the propagation path. For indoor office environments, $f=4$ is the default value \cite{WiMAXForum2008}; however, here, the small cell network is assumed to be deployed in a general domain including outdoor urban areas where there are less penetrated walls and floors. So, we use $f=3$ in our numerical study.

    It is assumed that the total transmission power including the antenna gain of each small cell base station is $P_{\mathrm{Tx}, [\textrm{dBm}]} = 32~\textrm{dBm}$. Shadowing is modeled as a random variable with lognormal
distribution with an underlying Gaussian distribution of zero mean and $8~\dB$ standard deviation.
The signal strength received at any distance $d$ from a base station $i$ is expressible as:
    \begin{equation*}
        P_{i, [\textrm{dBm}]}(d) = P_{\mathrm{Tx}, [\textrm{dBm}]} - PL_{[\dB]}(d) + X_i^{\dB}.
    \end{equation*}
    By (\ref{eq:PL_dB}), we have:
    \begin{multline}\label{eq:PdBm}
        P_{i, [\textrm{dBm}]}(d) = P_{\mathrm{Tx}, [\textrm{dBm}]}-37-18.3 f^{(\frac{f+2}{f+1}-0.46)} \\
        - 30\log_{10}(d) + X_i^{\dB}.
    \end{multline}

    The parameters $A$ and $\beta$ appearing in $P_i = A d^{-\beta} X_i$ can be identified from \eqref{eq:PdBm} after converting the received signal strength from the dBm scale to the linear scale:
    \begin{equation*}
        \beta = 3, \quad \textrm{and} \quad A_{[\textrm{mW}]} = 10^{\frac{P_{\mathrm{Tx}, [\textrm{dBm}]}-37-18.3 f^{(\frac{f+2}{f+1}-0.46)}}{10}}.
    \end{equation*}

    The received noise power $N_0$ is given by:
    \begin{equation*}
        N_{0, [\textrm{mW}]} = k_{\textrm{B}}T_{\textrm{Kelvin}} \times NF_{[\textrm{W}]} \times W_{[\textrm{Hz}]} \times 10^3,
    \end{equation*}
    where the effective bandwidth $W_{[\textrm{Hz}]} = 3.84 \times 10^6$ Hz, $k_{\textrm{B}}$ is the Boltzmann constant, and $T_{\textrm{Kelvin}}$ is the temperature in Kelvin,
$k_{\textrm{B}}T_{\textrm{Kelvin}} = 1.3804 \times 10^{-23} \times 290$ W/Hz and $NF_{[\dB]}$ is equal to 7~dB.

    It is assumed that the mobile is capable of scanning eight identified cells within 200~ms \cite{TS25.1332009}. So, the average time needed to scan one cell is given by $s = 25~\textrm{ms}$.

In order to check the accuracy of the approximations used in the analysis, a simulation was built with the above parameter setting. The interference field was generated according to a Poisson point process of intensity $\lambda$ in a region between $R_{\min}$ and $R_{\infty} = 100~\textrm{km}$. For a number $n$, the maximum of SINR received from $n$ base stations which are randomly selected from the disk $B$ between radii $R_{\min}$ and $R_B$ was computed. After that the expectation of the maximal capacity $\E\{\xi_n\}$ received from the $n$ selected BSs was evaluated.

In Fig.~\ref{fig:numresult}(a), the expectation of the maximal throughput $\E\{\xi_n\}$ for different $n$ is plotted, as obtained through the analytical model and simulation. The agreement between model and simulation is quite evident. As shown in Fig.~\ref{fig:numresult}(a), $\E\{\xi_n\}$ increases with $n$, though the increasing rate is slow down as $n$ increases. Note that in Fig.~\ref{fig:numresult}(a), $\E\{\xi_n\}$ is plotted after normalization by $\E\{\xi_{250}\}$.

Fig.~\ref{fig:numresult}(b) gives an example of acceleration $\rho_n$ for $T = 0.5$ second and
$L(n) = n \times 25~\textrm{ms}$. In the plot, $\rho_n$ is normalized by its maximum. Here, an agreement between model and simulation is also obtained. We see that $\rho_n$ first increases rapidly with $n$, attains its maximum at $n=42$ by simulation and $n = 43$ by model, and then decays.

    Next, using the model we compute the optimal number of cells to be scanned and the growth factor $g_n$ for different $T$. Note that in \eqref{eq:productivity}, the factor $T/(T + L(n))$ can be re-written as:
    \begin{equation*}
        \frac{T}{T + L(n)} = \frac{1}{1 + n \times s/T}
        \quad \textrm{for } \begin{cases} T < \infty, \\ L(n) = n\times s \end{cases}.
    \end{equation*}
    It is clear that this factor also depends on the ratio $T/s$. Fig.~\ref{fig:numresult}(c) plots the optimal $n$ for different values of $T/s$. Larger $T/s$ will drive the optimal $n$ towards larger values. Since $T$ can be roughly estimated
as the mobile residence time in a cell, which is proportional to the cell diameter divided by the
user speed, this can be rephrased by stating that the faster the mobile,
the smaller $T$ and thus the fewer cells the mobile should scan.

    Finally, Fig.~\ref{fig:numresult}(d) plots the growth factor $g_n$ with different $T$.
In Fig.~\ref{fig:numresult}(d), the ``limiting case'' corresponds to the case when $T \to \infty$ or $L(n) = 0$.
We see that $g_n$ is quite stable w.r.t. the  variation of $T$. Besides, $g_n$
flattens out at about 30 cells for a wide range of $T$. Therefore, in practice this value can be
taken as a recommended number of cells to be scanned in the system.

\section{Concluding Remarks}

    In this paper, we firstly develop asymptotic properties of the signal strength in cellular networks. We have shown that the signal strength received at the center of a ring shaped domain $B$ from a base station located in $B$ belongs to the maximum domain of attraction of a Gumbel distribution. Moreover, the maximum signal strength and the interference received from $n$ cells in $B$ are asymptotically independent as $n \to \infty$. The above properties are proved under the assumption that sites are uniformly distributed in $B$ and that shadowing is lognormal.
Secondly, the distribution of the best signal quality is derived.
These results are then used to optimize scanning in small cell networks.
We determine the number of cells to be scanned
for maximizing the mean user throughput within this setting.

\bibliographystyle{IEEEtran}
\bibliography{eurasip10_refs_revised}

\begin{thebibliography}{10}
\providecommand{\url}[1]{#1}
\csname url@samestyle\endcsname
\providecommand{\newblock}{\relax}
\providecommand{\bibinfo}[2]{#2}
\providecommand{\BIBentrySTDinterwordspacing}{\spaceskip=0pt\relax}
\providecommand{\BIBentryALTinterwordstretchfactor}{4}
\providecommand{\BIBentryALTinterwordspacing}{\spaceskip=\fontdimen2\font plus
\BIBentryALTinterwordstretchfactor\fontdimen3\font minus
  \fontdimen4\font\relax}
\providecommand{\BIBforeignlanguage}[2]{{%
\expandafter\ifx\csname l@#1\endcsname\relax
\typeout{** WARNING: IEEEtran.bst: No hyphenation pattern has been}%
\typeout{** loaded for the language `#1'. Using the pattern for}%
\typeout{** the default language instead.}%
\else
\language=\csname l@#1\endcsname
\fi
#2}}
\providecommand{\BIBdecl}{\relax}
\BIBdecl

\bibitem{Lee1991}
W.~C.~Y. Lee, ``Smaller cells for greater performance,'' \emph{IEEE
  Communications Magazine}, vol.~29, no.~11, pp. 19--23, Nov. 1991.

\bibitem{Claussen2007}
H.~Claussen, L.~T.~W. Ho, and L.~G. Samuel, ``Financial analysis of a
  pico-cellular home network deployment,'' in \emph{IEEE International
  Conference on Communications}, Jun. 2007, pp. 5604--5609.

\bibitem{Chandrasekhar2008}
V.~Chandrasekhar, J.~Andrews, and A.~Gatherer, ``Femtocell networks: a
  survey,'' \emph{IEEE Communications Magazine}, vol.~46, no.~9, pp. 59--67,
  Sep. 2008.

\bibitem{Saunders2009}
S.~Saunders, S.~Carlaw, A.~Giustina, R.~R. Bhat, V.~S. Rao, and R.~Siegberg,
  \emph{Femtocells: Opportunities and Challenges for Business and
  Technology}.\hskip 1em plus 0.5em minus 0.4em\relax Wiley, Jun. 2009.

\bibitem{URIE2009}
\BIBentryALTinterwordspacing
A.~Urie, ``Keynote: The future of mobile networking will be small cells,'' in
  \emph{IEEE International Workshop on Indoor and Outdoor Femto Cells}, Sep.
  2009. [Online]. Available: \url{http://iosc-workshop.homeip.net}
\BIBentrySTDinterwordspacing

\bibitem{Chika2009}
\BIBentryALTinterwordspacing
Y.~Chika, ``Keynote: True {BWA} - e{X}tended {G}lobal {P}latform,'' in
  \emph{IEEE International Workshop on Indoor and Outdoor Femto Cells}, Sep.
  2009. [Online]. Available: \url{http://iosc-workshop.homeip.net}
\BIBentrySTDinterwordspacing

\bibitem{Judge2009}
P.~Judge, ``{Vodafone launches home 3G Femtocell in the UK},'' {eWeekEurope},
  June 2009.

\bibitem{Leadbetter1983}
M.~R. Leadbetter, G.~Lindgren, and H.~Rootz{\'e}n, \emph{{Extremes and Related
  Properties of Random Sequences and Processes}}.\hskip 1em plus 0.5em minus
  0.4em\relax Springer Verlag, 1983.

\bibitem{Embrechts1997}
P.~Embrechts, C.~Kl\"{u}ppelberg, and T.~Mikosch, \emph{Modelling Extremal
  Events for Insurance and Finance}.\hskip 1em plus 0.5em minus 0.4em\relax
  Springer, Feb. 1997.

\bibitem{Nawrocki2006}
M.~Nawrocki, H.~Aghvami, and M.~Dohler, \emph{Understanding UMTS Radio Network
  Modelling, Planning and Automated Optimisation: Theory and Practice}.\hskip
  1em plus 0.5em minus 0.4em\relax John Wiley \& Sons, 2006.

\bibitem{WiMAX2007}
{WiMAX~Forum}, ``{Mobile System Profile},'' Approved Spec. Release 1.0,
  Revision 1.4.0, May 2007.

\bibitem{TS36.3312009}
{3GPP~TS~36.331}, ``{Evolved Universal Terrestrial Radio Access (E-UTRA) Radio
  Resource Control (RRC): Protocol Specification (Release 8)},'' Tech. Spec.
  v8.8.0, Dec. 2009.

\bibitem{NGMN2008}
{NGMN Alliance}, ``{Next Generation Mobile Networks} {U}se cases related to
  self-organising network, {O}verall description,'' Tech. Rep. v2.02, Dec.
  2008.

\bibitem{NGMN2008a}
------, ``{Next Generation Mobile Networks} {R}ecommendation on {SON} and
  {O\&M} requirements,'' Req. Spec. v1.23, Dec. 2008.

\bibitem{Magnusson1997}
S.~Magnusson and H.~Olofsson, ``Dynamic neighbor cell list planning in a
  microcellular network,'' in \emph{IEEE International Conference on Universal
  Personal Communications Record}, Oct. 1997, pp. 223--227.

\bibitem{Guerzoni2005}
R.~Guerzoni, I.~Ore, K.~Valkealahti, and D.~Soldani, ``Automatic neighbor cell
  list optimization for {UTRA FDD} networks: theoretical approach and
  experimental validation,'' \emph{WPMC, Aalborg, Denmark}, 2005.

\bibitem{Soldani2007}
D.~Soldani and I.~Ore, ``Self-optimizing neighbor cell list for {UTRA} {FDD}
  networks using detected set reporting,'' in \emph{IEEE 65th Vehicular
  Technology Conference}, 2007, pp. 694--698.

\bibitem{Amirijoo2008}
M.~Amirijoo, P.~Frenger, F.~Gunnarsson, H.~Kallin, J.~Moe, and K.~Zetterberg,
  ``Neighbor cell relation list and measured cell identity management in
  {LTE},'' in \emph{IEEE Network Operations and Management Symposium}, Apr.
  2008, pp. 152--159.

\bibitem{Moe2009}
M.~Z. Win, P.~C. Pinto, and L.~A. Shepp, ``A mathematical theory of network
  interference and its applications,'' \emph{Proceedings of the IEEE}, vol.~97,
  no.~2, pp. 205--230, Feb. 2009.

\bibitem{Baccelli2009}
\BIBentryALTinterwordspacing
F.~{B}accelli and B.~B{\l}aszczyszyn,
  \emph{\BIBforeignlanguage{{A}nglais}{{S}tochastic {G}eometry and {W}ireless
  {N}etworks, {V}olume {I} - {T}heory}}, 2009, vol.~1. [Online]. Available:
  \url{http://hal.inria.fr/inria-00403039/en/}
\BIBentrySTDinterwordspacing

\bibitem{TR36.9422009}
{3GPP~TR~36.942}, ``{Evolved Universal Terrestrial Radio Access (E-UTRA): Radio
  Frequency (RF) system scenarios (Release 8)},'' Tech. Rep. v8.2.0, May 2009.

\bibitem{WiMAXForum2008}
{WiMAX~Forum}, ``{WiMAX} systems evaluation methodology,'' Spec. v2.1, Jul.
  2008.

\bibitem{Chow1978}
T.~Chow and J.~Teugels, ``The sum and the maximum of i.i.d. random variables,''
  in \emph{Proc. Second Prague Symp. Asymptotic Statistics}, 1978, pp. 81--92.

\bibitem{Anderson1991}
C.~W. Anderson and K.~F. Turkman, ``The joint limiting distribution of sums and
  maxima of stationary sequences,'' \emph{Journal of Applied Probability},
  vol.~28, no.~1, pp. 33--44, 1991.

\bibitem{IEEE802.162009}
{IEEE~802.16}, ``{Air Interface for Broadband Wireless Access Systems},'' IEEE,
  Standard Std 802.16-2009, May 2009.

\bibitem{TS36.3002009}
{3GPP~TS~36.300}, ``{Evolved Universal Terrestrial Radio Access (E-UTRA) and
  Evolved Universal Terrestrial Radio Access Network (E-UTRAN) - Overall
  description: Stage 2 (Release 8)},'' Tech. Spec. v8.11.0, Dec. 2009.

\bibitem{TS25.3312009}
{3GPP~TS~25.331}, ``{Radio Resource Control (RRC): Protocol Specification
  (Release 8)},'' Tech. Spec. v8.6.0, Mar. 2009.

\bibitem{TS25.1332009}
{3GPP~TS~25.133}, ``{Requirements for support of Radio Resource Management FDD
  (Release 8)},'' Tech. Spec. v8.9.0, Dec. 2009.

\bibitem{TR101.1121998}
{ETSI~TR~101.112}, ``{Selection procedures for the choice of radio transmission
  technologies of the UMTS},'' Tech. Rep. v3.2.0, Apr. 1998.

\bibitem{ABRAMOWITZ1972}
M.~Abramowitz and I.~A. Stegun, \emph{Handbook of Mathematical
  Functions}.\hskip 1em plus 0.5em minus 0.4em\relax Dover Publications, 1965.

\bibitem{Takahashi1987}
R.~Takahashi, ``Normalizing constants of a distribution which belongs to the
  domain of attraction of the {Gumbel} distribution,'' \emph{Statistics \&
  Probability Letters}, vol.~5, no.~3, pp. 197--200, 1987.

\bibitem{Feller1971}
W.~Feller, \emph{An Introduction to Probability Theory and Its Applications},
  2nd~ed.\hskip 1em plus 0.5em minus 0.4em\relax John Wiley \& Sons, 1971,
  vol.~2.

\end{thebibliography}

\appendices

\section{Proof of Lemma~\ref{lem:cdfP}} \label{appdA}

        Let $d_i = |\bfy - \bfx_i|$ be the distance from a site located at $\bfx_i \in \R^2$ to a position $\bfy \in \R^2$. Under the assumption that site locations are uniformly distributed in $B$, the distance $d_i$ from a site located in $B$, i.e., $\bfx_i \in B$, to the center of $B$ has the following distribution:
        \begin{equation*}
            F_D(d) = \Pb{[d_i \leq d]} = \frac{\pi d^2 - \pi R_{\min}^2}{\pi R_B^2 - \pi R_{\min}^2} = \frac{d^2 - R_{\min}^2}{R_{B}^2 - R_{\min}^2}.
        \end{equation*}

        Let $U_i = A d_i^{-\beta}$, for $\beta > 0$, its distribution is equal to:
        \begin{equation*}
            F_U(u) = -c\bigr( u^{-\frac{2}{\beta}} - a^{-\frac{2}{\beta}} \bigr), \quad \mathrm{for} \quad u \in [a,b],
        \end{equation*}
        where $c = A^{\frac{2}{\beta}} (R_{B}^2 - R_{\min}^2)^{-1} $, $a = A R_{B}^{-\beta}$, and $b = A R_{\min}^{-\beta}$. The density of $U_i$ is given by $f_U(u) = (2c/\beta) u^{-1-2/\beta}$.

        Thus, the distribution $F_P$ of the power $P_i$ is equal to:
        \begin{equation}\label{eq:power1}
            F_P(x) = \int_{u=a}^b {F_X(\frac{x}{u}) f_U(u) \di u}.
        \end{equation}

        Substituting $F_X$ with lognormal distribution of parameters $(0, \sigma_X)$ and $f_U$ given above into (\ref{eq:power1}), after changing the variable such that $t = \log (\frac{x}{u})$, we have:
        \begin{multline*}
            F_P(x) = \frac{c}{\beta} x^{-\frac{2}{\beta}} \Big( \int_{\log {(\frac{x}{b})}}^{\log {(\frac{x}{a})}} {e^{\frac{2t}{\beta}} \di t} \\
            + \int_{\log {(\frac{x}{b})}}^{\log {(\frac{x}{a})}} {e^{\frac{2t}{\beta}} \erf \bigr( \frac{t}{\sqrt{2}\sigma_X}\bigr) \di t} \Big)
        \end{multline*}
        where the first integral is straightforward. By doing an integration by parts of $\erf\bigr( \frac{t}{\sqrt{2}\sigma_X}\bigr)$ and $e^{\frac{2t}{\beta}} \di t$ for the second integral, we get:
        \begin{multline*}
            F_P(x) = \frac{c}{\beta} x^{-\frac{2}{\beta}} \frac{\beta}{2} \Big[e^{\frac{2t}{\beta}} + e^{\frac{2t}{\beta}}\erf \bigr( \frac{t}{\sqrt{2}\sigma_X}\bigr) \\
            - e^{\frac{2\sigma_X^2}{\beta^2}} \erf\bigr(\frac{t}{\sqrt{2}\sigma_X} - \frac{\sqrt{2}\sigma_X}{\beta} \bigr) \Big] \Big|_{t=\log {(\frac{x}{b})}}^{\log {(\frac{x}{a})}}.
        \end{multline*}
        After some elementary simplifications, we can obtain:
        \begin{multline*}
            F_P(x) = c \Big\{ a^{-\frac{2}{\beta}} \Big(\frac{1}{2} + \frac{1}{2} \erf(\frac{\log x - \mu_1}{\sqrt{2}\sigma_X}) \Big) \\
            - b^{-\frac{2}{\beta}} \Big(\frac{1}{2} + \frac{1}{2} \erf(\frac{\log x - \mu_2}{\sqrt{2}\sigma_X}) \Big) + e^\nu x^{-\frac{2}{\beta}} \Big[ -\frac{1}{2} - \frac{1}{2} \erf(\frac{\log x - \mu_3}{\sqrt{2}\sigma_X}) \\
            + \frac{1}{2} + \frac{1}{2} \erf(\frac{\log x - \mu_4}{\sqrt{2}\sigma_X}) \Big] \Big\}
        \end{multline*}
        where $\nu = \frac{2\sigma_X^2}{\beta^2}$, $\mu_1 = \log a$, $\mu_3 = \mu_1 + 2\sigma_X^2/\beta$, $\mu_2 = \log b$, and $\mu_4 = \mu_2 + 2\sigma_X^2/\beta$. Let $G_j, \, j = 1,...,4$, be the lognormal distribution of parameters $(\mu_j, \sigma_X), \, j = 1,...,4$, $F_P$ can be rewritten as (\ref{eq:cdfPi}). \hfill $\square$

\section{Proof of Lemma~\ref{lem:tailequivP}}\label{appdB}

        Let $\bar{G}_j(x) = 1 - G_j(x)$ and note that $c(a^{-\frac{2}{\beta}} - b^{-\frac{2}{\beta}}) = 1$, we have from (\ref{eq:cdfPi}):
        \begin{multline*}
            F_P(x) = 1 - c \Big\{ a^{-\frac{2}{\beta}}\bar{G}_1(x) - b^{-\frac{2}{\beta}}\bar{G}_2(x) \\
            - e^{\nu}x^{-\frac{2}{\beta}}\bar{G}_3(x) + e^{\nu}x^{-\frac{2}{\beta}}\bar{G}_4(x) \Big\}.
        \end{multline*}
        This yields the tail distribution $\bar{F}_p = 1 - F_P$:
        \begin{multline}\label{eq:appdx_tailP1}
            \bar{F}_p(x) = c \Big\{ a^{-\frac{2}{\beta}}\bar{G}_1(x) - b^{-\frac{2}{\beta}}\bar{G}_2(x) \\
            - e^{\nu}x^{-\frac{2}{\beta}}\bar{G}_3(x) + e^{\nu}x^{-\frac{2}{\beta}}\bar{G}_4(x) \Big\}.
        \end{multline}

        For (\ref{eq:appdx_tailP1}), we have $\bar{G}_j(x) = \frac{1}{2}\, \erfc\{(\log x - \mu_j)/(\sqrt{2}\sigma_X)\}$. An asymptotic expansion of $\erfc(x)$ for large $x$ \cite[7.1.23]{ABRAMOWITZ1972} gives us:
        \begin{multline*}
            \bar{G}_3(x) \underset{\infty}{\thicksim} \frac{\sigma_X}{\sqrt{2\pi} (\log x - \mu_3)} \exp\Big\{-\frac{(\log x - \mu_3)^2}{2\sigma_X^2} \Big\} \\
            = \frac{\sigma_X}{\sqrt{2\pi} (\log x - \mu_1 - \frac{2\sigma_X^2}{\beta})} \exp\Big\{-\frac{(\log x - \mu_1 - \frac{2\sigma_X^2}{\beta})^2}{2\sigma_X^2} \Big\} \\
            = \frac{\sigma_X a^{-\frac{2}{\beta}} e^{-\nu}x^{\frac{2}{\beta}}}{\sqrt{2\pi}(\log x - \mu_1)}\exp\Big\{-\frac{(\log x - \mu_1)^2}{2\sigma_X^2} \Big\} \frac{1}{1 - \frac{2\sigma_X^2}{\beta (\log x - \mu_1)}}
        \end{multline*}
        in which after a Taylor expansion of the last term on the right-hand side, we can have:
        \begin{displaymath}
            e^{\nu}x^{-\frac{2}{\beta}} \bar{G}_3(x) \underset{\infty}{\thicksim} a^{-\frac{2}{\beta}} \Bigr[\bar{G}_1(x) + \sqrt{\frac{2}{\pi}}\frac{\sigma_X^3}{\beta} \frac{\exp\big(-\frac{(\log x - \mu_1)^{2}}{2\sigma_X^2}\big)}{(\log x - \mu_1)^2} \Bigr].
        \end{displaymath}
        This implies that
        \begin{multline}\label{eq:G3toG1}
            a^{-\frac{2}{\beta}} \bar{G}_1(x) - e^{\nu} x^{-\frac{2}{\beta}} \bar{G}_3(x) \\ \underset{\infty}{\thicksim} - \sqrt{\frac{2}{\pi}}\frac{a^{-\frac{2}{\beta}}\sigma_X^3}{\beta} \frac{\exp\big(-\frac{(\log x - \mu_1)^2}{2\sigma_X^2}\big)}{(\log x - \mu_1)^2}.
        \end{multline}

        In the same manner, we have
        \begin{multline}\label{eq:G4toG2}
            b^{-\frac{2}{\beta}} \bar{G}_2(x) - e^{\nu} x^{-\frac{2}{\beta}} \bar{G}_4(x)\\ \underset{\infty}{\thicksim} - \sqrt{\frac{2}{\pi}}\frac{b^{-\frac{2}{\beta}}\sigma_X^3}{\beta} \frac{\exp\big(-\frac{(\log x - \mu_2)^2}{2\sigma_X^2}\big)}{(\log x - \mu_2)^2}.
        \end{multline}

        A substitution of (\ref{eq:G3toG1}) and (\ref{eq:G4toG2}) into (\ref{eq:appdx_tailP1}) results in
        \begin{multline}\label{eq:power5}
            \bar{F}_p(x) \underset{\infty}{\thicksim} \sqrt{\frac{2}{\pi}}\frac{c\sigma_X^3}{\beta} \Big\{b^{-\frac{2}{\beta}}\frac{\exp\big(-(\log x - \mu_2)^2/(2\sigma_X^2)\big)}{(\log x - \mu_2)^2} \\
            - a^{-\frac{2}{\beta}}\frac{\exp\big(-(\log x - \mu_1)^2/(2\sigma_X^2)\big)}{(\log x - \mu_1)^2} \Big\}.
        \end{multline}

        Moreover, $b > a$ yields $\mu_2 - \mu_1 = \log(b/a) > 0$. Then, we have the following result for large $x$:
        \begin{multline*}
            \frac{\exp\big( - \frac{(\log x - \mu_1)^2}{2\sigma_X^2}\big) / (\log x - \mu_1)^{2}}{\exp\big(-\frac{(\log x - \mu_2)^2}{2\sigma_X^2}\big)/ (\log x - \mu_2)^{2}}\\
            = \Big(\frac{\log x - \mu_2}{\log x - \mu_1}\Big)^{2} \exp\Big(\frac{\mu_2^2 - \mu_1^2}{2\sigma_X^2} \Big) x^{-\frac{\mu_2 - \mu_1}{\sigma_X^2}} \underset{x \to \infty}{\to} 0.
        \end{multline*}
        Taking this into account in (\ref{eq:power5}), finally we have:
        \begin{subequations}
            \begin{align}
            \bar{F}_p(x) & \underset{\infty}{\thicksim} \kappa \frac{\exp\big(-(\log x - \mu_2)^2/(2\sigma_X^2)\big)}{\big((\log x - \mu_2)/(\sqrt{2}\sigma_X)\big)^2} \nonumber \\ 
            & \underset{\infty}{\thicksim} 2\sqrt{2\pi}\sigma_X\kappa\frac{\bar{G}_2(x)}{\log x - \mu_2}, \quad \kappa = \frac{\sigma_X}{\sqrt{2\pi}\beta}\frac{R_{\min}^2}{R_{B}^2-R_{\min}^2}.\nonumber 
            \end{align}
        \end{subequations}
         \hfill $\square$

\section{Proof of Theorem~\ref{thm:mdaPi}}\label{appdC}

    We will use Lemma~\ref{lem:tailequivP} and the following two lemmas to prove Theorem~\ref{thm:mdaPi}.

    \begin{lem}[Embrechts et al.~\cite{Embrechts1997}]\label{lem:monotrans}
        \emph{Let $Z_i$ be i.i.d. random variables having distribution $F$, and $\Psi_n = \max_{i=1}^n Z_i$. Let $g$ be an increasing real function, denote $\tilde{Z}_i = g(Z_i)$, and $\tilde{\Psi}_n = \max_{i=1}^n \tilde{Z}_i$. If $F \in \textrm{MDA}(\Lambda)$ with normalizing constant $c_n$ and $d_n$, then
        \begin{equation*}
            \lim_{n \to \infty} {\Pb\big(\tilde{\Psi}_n \leq g(c_n z + d_n) \big)} = \Lambda(z), \quad z \in \R.
        \end{equation*}}
    \end{lem}

    \begin{lem}[Takahashi~\cite{Takahashi1987}]\label{lem:TakProp1}
        \emph{Let $F$ be a distribution function. Suppose that there exists constants $\omega > 0$, $l > 0$, $\eta > 0$ and $r \in \R$ such that
        \begin{equation}\label{eq:TakProp1}
            \lim_{x \to \infty}{\bigr(1 - F(x)\bigr)/\bigr(l x^r e^{-\eta x^{\omega}}\bigr)} = 1.
        \end{equation}
        For $\mu \in \R$ and $\sigma > 0$, let $F_{\ast} = F((x - \mu)/\sigma)$. Then, $F_{\ast} \in \textrm{MDA}(\Lambda)$ with normalizing constants $c_n^{\ast} = \sigma c_n$ and $d_n^{\ast} = \sigma d_n + \mu$, where
        \begin{eqnarray*}
            c_n & = & \frac{(\log{n}/\eta)^{\frac{1}{\omega}-1}}{\omega \eta}, \; \textrm{and} \\
            d_n & = & \big(\frac{\log n}{\eta}\big)^{1/\omega} + \frac{\eta^{1/\omega}}{\omega^2} \frac{r(\log{\log n} - \log \eta) + \omega \log{l}}{(\log n)^{1 - \frac{1}{\omega}}}.
        \end{eqnarray*}}
    \end{lem}

    Let $g(t) = e^{\sqrt{2}\sigma_X t + \mu_2}$ be a real function defined on $\R$, $g$ is increasing with $t$. Let $Q_i$ be the random variable such that $P_i = g(Q_i)$. By \eqref{eq:tailequivPowerA} of Lemma~\ref{lem:tailequivP}, the tail distribution $\bar{F}_Q$ is given by:
    \begin{equation}
        \bar{F}_Q(x) = \bar{F}_P\big(e^{\sqrt{2}\sigma_X x + \mu_2}\big) \thicksim \kappa x^{-2} e^{-x^2}, \quad \textrm{as } x \to \infty. \label{eq:Fqbar}
    \end{equation}

    By \eqref{eq:Fqbar}, $F_Q$ satisfies Lemma~\ref{lem:TakProp1} with constants $l = \kappa$, $r = -2$, $\eta = 1$, and $\omega = 2$. So, $F_{Q} \in \textrm{MDA}(\Lambda)$ with the following normalizing constants:
    \begin{equation}\label{eq:normconstofQ}
    \begin{split}
        c_n^{\ast} = & \frac{(\log{n}/\eta)^{\frac{1}{\omega}-1}}{\omega \eta} = \frac{1}{2}(\log n)^{-\frac{1}{2}}, \qquad \textrm{and} \\
        d_n^{\ast} = & \big(\frac{\log n}{\eta}\big)^{1/\omega} + \frac{\eta^{1/\omega}}{\omega^2} \frac{r(\log{\log n} - \log \eta) + \omega \log{l}}{(\log n)^{1 - \frac{1}{\omega}}} \\
        = & (\log n)^{\frac{1}{2}} + \frac{1}{2} \frac{(-\log{\log n} + \log{\kappa})}{(\log n)^{\frac{1}{2}}}.
    \end{split}
    \end{equation}

    Then, by Lemma~\ref{lem:monotrans}, we have
    \begin{eqnarray*}
        \lim_{n \to \infty} {\Pb\Big(M_n \leq g(c_n^{\ast} x + d_n^{\ast}) \Big)} = \Lambda(x), \quad x \in \R.
    \end{eqnarray*}
    By a Taylor expansion of $\exp(\sqrt{2}\sigma_X c_n^{\ast}x)$, we have:
    \begin{multline*}
        \lim_{n \to \infty} {\Pb\big\{e^{-(\sqrt{2}\sigma_X d_n^{\ast} + \mu_2)} M_n \leq 1 + \sqrt{2}\sigma_X c_n^{\ast}x + o(c_n^{\ast})\big\}} \\
        = \Lambda(x).
    \end{multline*}
    Since $c_n^{\ast} \to 0$ when $n \to \infty$, we have
    \begin{equation}\label{eq:limPsin}
        \frac{M_n - e^{\sqrt{2}\sigma_X d_n^{\ast} + \mu_2}}{\sqrt{2}\sigma_X c_n^{\ast} e^{\sqrt{2}\sigma_X d_n^{\ast} + \mu_2}} \overset{d}{\to} \Lambda, \qquad \textrm{as } n \to \infty.
    \end{equation}

    Substituting $c_n^{\ast}$ and $d_n^{\ast}$ from (\ref{eq:normconstofQ}) into (\ref{eq:limPsin}), we obtain $c_n$ and $d_n$ for (\ref{eq:maxnormconstants}). The conditions $R_{\max} < \infty$, $R_{\min} > 0$ and $\sigma_X > 0$ provide $\kappa > 0$. This leads to $d_n > 0$, and consequently, $c_n > 0$.  \hfill $\square$

\section{Proof of Lemma~\ref{lem:cfI}}\label{appd:lem:cfI}

    Under the assumptions of the lemma,
the interference field can be modeled as a shot noise defined on $\R^2$ excluding the inner disk of radius $R_{\min}$. Hence, using Proposition 2.2.4 in \cite{Baccelli2009}, the Laplace transform of $I$ is given by:
    \begin{equation}\label{eq:laplaceI}
        \mathcal{L}_I(s) = \exp\Big\{-2\pi\lambda \int_{R_{\min}}^{\infty}{\Big(1 - \E\{e^{-\frac{s A X_i}{r^{\beta}}}\}\Big) r \di r}\Big\}.
    \end{equation}
    Noting that
    \begin{equation*}
        \phi_I(w) = \mathcal{L}_I(- j w), \quad w \in \R,
    \end{equation*}
    we have from \eqref{eq:laplaceI} that:
    \begin{multline}\label{eq:cfI:0}
        \phi_I(w) = \exp\Big\{-2\pi\lambda \int_{R_{\min}}^{\infty}(1 - \E\{e^{\frac{j w A X_i}{r^{\beta}}}\})r \di r\Big\}.
    \end{multline}
    Using the change of variable $t = |w| A r^{-\beta}$, we obtain
    \begin{multline}
        \int_{R_{\min}}^{+\infty}(1 - \E\{e^{\frac{j w A X_i}{r^{\beta}}}\})r \di r\\
        = \frac{(A |w|)^{2/\beta}}{\beta} \int_0^{\frac{A |w|}{R_{\min}^{\beta}}}\frac{1 - \E\{e^{j \sign(w) t X_i}\}}{t^{2/\beta + 1}} \di t,
    \end{multline}
    where $\E\{e^{j \sign(w) t X_i}\} = \phi_X(\sign(w)t)$. So, substituting this into \eqref{eq:cfI:0}, we get the first part of the Lemma~\ref{lem:cfI}.

    From \eqref{eq:cfI:exact}, for all $p = 1, 2, \ldots$, we have:
    \begin{multline}
        |\phi_I(w)|^p = \exp\Big( \\
        - p \pi\lambda\alpha (A|w|)^{\alpha} \E\big\{\int_0^{\frac{A|w|}{R_{\min}^{\beta}}}\frac{1 - \cos(t X_i)}{t^{\alpha + 1}} \di t \big\}\Big).
    \end{multline}
    Since $1 - \cos(t X_i) \ge 0$, $\forall t \in \R$, we have
    \begin{equation}
        \E\Big\{\int_0^{\frac{A|w|}{R_{\min}^{\beta}}}\frac{1 - \cos(t X_i)}{t^{\alpha + 1}} \di t \Big\} \ge 0.
    \end{equation}
    Therefore
    \begin{equation}
        |\phi_I(w)|^p \le \exp(- c |w|^{\alpha}),
    \end{equation}
    where $c$ is some positive constant, and hence the right hand-side of this is an absolutely integrable function. This proves the second assertion of Lemma~\ref{lem:cfI}.

    Under the assumption that $A R_{\min}^{-\beta} \approx \infty$, $\phi_I$ can be approximated by:
    \begin{equation}\label{eq:cfI:approx:proof}
        \phi_I(w) \approx \exp\bigr(- \pi\lambda\alpha(A|w|)^{\alpha} \int_0^{\infty}\frac{1 - \phi_X(\sign(w)t)}{t^{\alpha + 1}} \di t\bigr).
    \end{equation}
    For $0 < \alpha < 1$, we have
    \begin{multline}
        \int_0^{\infty}\frac{1 - e^{j \sign(w) t X_i}}{t^{\alpha + 1}} \di t \\
        = -\Gamma(-\alpha)|\sign(w)X_i|^{\alpha} e^{-j \sign(w)\frac{\pi\alpha}{2}}
    \end{multline}
    Since $X_i \ge 0$, we can write $|\sign(w)X_i|^{\alpha} = X_i^{\alpha}$.
Taking expectations on both sides, we get
    \begin{align*}
        \int_0^{+\infty} & \frac{1 -\E\{e^{j \sign(w) t X_i}\}}{t^{\alpha + 1}} \di t \\
        & = - \E\{X_i^{\alpha}\} \Gamma(-\alpha) e^{-j \sign(w)\frac{\pi\alpha}{2}} \\
        & = \E\{X_i^{\alpha}\}  \frac{\Gamma(1-\alpha)}{\alpha} \cos(\frac{\pi\alpha}{2}) [1 - j \sign(w)\tan(\frac{\pi\alpha}{2})].
    \end{align*}
    Substituting this into \eqref{eq:cfI:approx:proof} and noting that
    \begin{equation}\label{eq:EXialpha}
        \E\{X_i^{\alpha}\} = \exp(\frac{1}{2}\alpha^2 \sigma_X^2),
    \end{equation}
    for $X_i$ lognormally distributed, we obtain \eqref{eq:cfI:approx}. \hfill $\square$

\section{Proof of Theorem~\ref{thm:cdfMaxSinrDenseNwk}}\label{appdD}

    Under the assumption that sites are distributed as a homogeneous Poisson point process of intensity $\lambda$ in $B$, the expected number of cells in $B$ is $\pi \lambda (R_{B}^2 - R_{\min}^2)$.
We assume that $\pi \lambda (R_{B}^2 - R_{\min}^2)$ is much larger than $n$, which ensures that
there are $n$ cells in $B$ with high probability, so that $Y_n$ is well defined.

    Under the conditions of Theorem~\ref{thm:cdfMaxSinrDenseNwk}, $M_n$ and $I$ are asymptotically independent according to Corollary~\ref{cor:asympindMnI}. So, by substituting \eqref{eq:asympjointIandM} into \eqref{eq:cdfMaxSinr1}, we have:
    \begin{equation}\label{eq:proofCdfMaxSinrDenseNwk:1}
        \Pb\{Y_n \geq \gamma\} \approx \int_{0}^{\infty} f_{M_n}(u)\!\! \int_{0}^{\infty}{\!\! h(v,u) f_{I}(v) \di v}  \di u
    \end{equation}
    where
        \begin{IEEEeqnarray}{rCl}
        h(v,u) & = & 1_{(v \leq \frac{(1+\gamma)u}{\gamma} - 1)} 1_{(v \geq u)} \nonumber\\
        & = & \begin{cases} 1 & \textrm{if } v \in [u,\frac{1+\gamma}{\gamma}u - 1] \textrm{ and } u > \gamma\\
        0 & \textrm{otherwise}\end{cases}.\label{eq:h}
    \end{IEEEeqnarray}

    It is easily seen that $h(v,u)$ is square-integrable with respect to $v$, and its Fourier transform w.r.t. $v$ is given by:
    \begin{equation}\label{eq:proofCdfMaxSinrDenseNwk:3}
        \hat{h}(\frac{w}{2\pi},u) = \begin{cases} 0 & \textrm{if } u \le \gamma \\ \int_{u}^{\frac{1+\gamma}{\gamma}u - 1}e^{-j w v}\di v & \textrm{if } u > \gamma \end{cases}
    \end{equation}
    which yields:
    \begin{equation}\label{eq:fourierofh}
        \hat{h}(\frac{w}{2\pi},u) = \begin{cases} 0 & \textrm{if } u \le \gamma \\ \frac{1}{jw}\Big(e^{-jwu} - e^{jw \big(1 \!-\! \frac{1+\gamma}{\gamma}u \big)}\Big) & \textrm{if } u > \gamma \end{cases}.
    \end{equation}

    Besides, according to Lemma~\ref{lem:cfI} we have that $\phi_I \in \mathbb{L}$ and $\phi_I \in \mathbb{L}^2$, where $\mathbb{L}^2$ is the space of square integrable functions. And so, by Theorem 3 in \cite[p.509]{Feller1971}, $f_I$ is bounded continuous and square integrable. Hence, by applying the Plancherel-Parseval theorem to the inner integral of
\eqref{eq:proofCdfMaxSinrDenseNwk:1}, we have
    \begin{equation}\label{eq:proofCdfMaxSinrDenseNwk:2}
        \int_{0}^{\infty}{h(v,u) f_{I}(v) \di v} = \int_{-\infty}^{\infty}{\hat{h}(-w,u) \hat{f}_{I}(w) \di w},
    \end{equation}
    where $\hat{f}_{I}(w)$ is the Fourier transform of $f_{I}(v)$. Take \eqref{eq:fourierofh} into account for \eqref{eq:proofCdfMaxSinrDenseNwk:2} and \eqref{eq:proofCdfMaxSinrDenseNwk:1}, we have:
    \begin{equation}\label{eq:proofCdfMaxSinrDenseNwk:4}
            \bar{F}_{Y_n}(\gamma) = \int_{\gamma}^{\infty}\!\!\!\Big\{f_{M_n}(u) \!\!\!
            \int_{-\infty}^{\infty}{\!\!\hat{h}(-w,u) \hat{f}_{I}(w) \di w}\Big\}\di u,
    \end{equation}
    where we further have
    \begin{IEEEeqnarray}{lll}
            \int_{-\infty}^{\infty}{\!\!\hat{h}(-w,u) \hat{f}_{I}(w) \di w} \nonumber \\
            = \frac{1}{2\pi}\int_{-\infty}^{+\infty}{\hat{h}(-\frac{w}{2\pi},u) \hat{f}_{I}(\frac{w}{2\pi}) \di w} \nonumber\\
            = \frac{1}{2\pi}\int_{0}^{+\infty}\!\!\!\!\!\!\!\! \Big(\hat{h}(\frac{w}{2\pi},u) \hat{f}_{I}(\frac{-w}{2\pi}) + \hat{h}(\frac{-w}{2\pi},u) \hat{f}_{I}(\frac{w}{2\pi})\Big)\di w.
            \label{eq:proofCdfMaxSinrDenseNwk:5}
    \end{IEEEeqnarray}
    Note that
    \begin{equation}
        \hat{f}_I(\frac{w}{2\pi}) = \phi_I(-w).
    \end{equation}
    And under the assumption that $A R_{\min}^{-\beta} \approx \infty$, $\phi_I$ is approximated by \eqref{eq:cfI:approx}. Thus, by \eqref{eq:cfI:approx} and \eqref{eq:fourierofh}, we have for $w \in [0, +\infty)$:
    \begin{multline}\label{eq:proofCdfMaxSinrDenseNwk:6}
            \hat{h}(\frac{w}{2\pi},u) \hat{f}_{I}(-\frac{w}{2\pi}) \approx \frac{e^{-\delta w^{\alpha}}}{j w}\Big\{e^{j(-w u + \delta w^{\alpha}\tan\frac{\pi\alpha}{2})} \\
            - e^{-j(-w+w\frac{1+\gamma}{\gamma}u - \delta w^{\alpha}\tan\frac{\pi\alpha}{2})}\Big\}
    \end{multline}
    and
    \begin{multline}\label{eq:proofCdfMaxSinrDenseNwk:7}
            \hat{h}(-\frac{w}{2\pi},u) \hat{f}_{I}(\frac{w}{2\pi}) \approx \frac{e^{-\delta w^{\alpha}}}{j w}\Big\{-e^{-j(-w u + \delta w^{\alpha}\tan\frac{\pi\alpha}{2})} \\
            + e^{j(-w+w\frac{1+\gamma}{\gamma}u - \delta w^{\alpha}\tan\frac{\pi\alpha}{2})}\Big\}.
    \end{multline}
    By \eqref{eq:proofCdfMaxSinrDenseNwk:6} and \eqref{eq:proofCdfMaxSinrDenseNwk:7}, we get
    \begin{IEEEeqnarray*}{lll}
            \!\!\!\!\!\!\!\!\!\!\!\!\!\!\!\! \frac{1}{2\pi}\Big[\hat{h}(\frac{w}{2\pi},u) \hat{f}_{I}(\frac{-w}{2\pi}) + \hat{h}(\frac{-w}{2\pi},u) \hat{f}_{I}(\frac{w}{2\pi})\Big]\\
            \approx \frac{e^{-\delta w^{\alpha}}}{\pi w}\Big[\sin(-w u + \delta w^{\alpha}\tan\frac{\pi\alpha}{2}) \\
            \qquad + \sin(-w+w\frac{1+\gamma}{\gamma}u - \delta w^{\alpha}\tan\frac{\pi\alpha}{2}) \Big] \\
            = \frac{2 e^{-\delta w^{\alpha}}}{\pi w} \sin\big(w\frac{u-\gamma}{2\gamma}\big) \\
            \qquad \times \cos\big(w u + w\frac{u-\gamma}{2\gamma} - \delta w^{\alpha}\tan{\frac{\pi \alpha}{2}}\big).
    \end{IEEEeqnarray*}

    Substitute the above into \eqref{eq:proofCdfMaxSinrDenseNwk:5} and then into \eqref{eq:proofCdfMaxSinrDenseNwk:4}, we have   \eqref{eq:cdfMaxSinrDenseNwk}. \hfill $\square$

\end{document}